\documentclass[aps,prl,superscriptaddress,11pt]{revtex4}

\usepackage[T1]{fontenc}
\usepackage{times}
\usepackage{color,graphicx}
\usepackage{subfig,array}
\usepackage[dvipdfmx,colorlinks=true]{hyperref}
\usepackage{amsthm,amssymb,amsmath}
\usepackage[colorlinks=true]{hyperref}

\newcommand\tr{\mathop{\rm tr}\nolimits}

\newtheorem{definition}{Definition}
\newtheorem{lemma}{Lemma}
\newtheorem{theorem}{Theorem}

\newtheorem{remark}{Remark}

\newcommand{\nc}{\newcommand}
\nc{\cA}{{\cal A}} \nc{\cB}{{\cal B}} \nc{\cC}{{\cal C}}
\nc{\cD}{{\cal D}} \nc{\cE}{{\cal E}} \nc{\cF}{{\cal F}}
\nc{\cG}{{\cal G}} \nc{\cH}{{\cal H}} \nc{\cI}{{\cal I}}
\nc{\cJ}{{\cal J}} \nc{\cK}{{\cal K}} \nc{\cL}{{\cal L}}
\nc{\cM}{{\cal M}} \nc{\cN}{{\cal N}} \nc{\cO}{{\cal O}}
\nc{\cP}{{\cal P}} \nc{\cQ}{{\cal Q}} \nc{\cR}{{\cal R}}
\nc{\cS}{{\cal S}} \nc{\cT}{{\cal T}} \nc{\cU}{{\cal U}}
\nc{\cV}{{\cal V}} \nc{\cW}{{\cal W}} \nc{\cX}{{\cal X}}
\nc{\cZ}{{\cal Z}}
\begin{document}

\title[Fidelity based measurement-induced quantum correlation]
{Fidelity based unitary operation-induced quantum correlation for
continuous-variable systems}

\author{Liang Liu}
\affiliation{Institute of Mechanics, Taiyuan University of
Technology, Taiyuan 030024, P. R.
China}\email{liuliang0107@link.tyut.edu.cn}
\author{Xiaofei Qi}
\affiliation{Department of Mathematics, Shanxi University, Taiyuan
030006, P. R. China} \affiliation{Institute of Big Data Science and
Industry, Shanxi University, Taiyuan 030024, P. R. China}
\email{xiaofeiqisxu@aliyun.com}
\author{Jinchuan Hou}
\affiliation{Department of Mathematics, Taiyuan University of
Technology, Taiyuan 030024, P. R. China}
 \email{houjinchuan@tyut.edu.cn, jinchuanhou@aliyun.com}

\begin{abstract}
We propose a measure of nonclassical correlation $N_{\mathcal
F}^{\mathcal G}$ in terms of local Gaussian unitary operations based
on square of the fidelity $\mathcal F$ for bipartite
continuous-variable systems. This quantity is easier to calculate or
estimate and is a remedy for the local ancilla problem associated
with the geometric measurement-induced nonlocality. A simple
computation formula of $N_{\mathcal F}^{\mathcal G}$ for any
$(1+1)$-mode Gaussian states is presented and an estimation of
$N_{\mathcal F}^{\mathcal G}$ for any $(n+m)$-mode Gaussian states
is given. For any $(1+1)$-mode Gaussian states,  $N_{\mathcal
F}^{\mathcal G}$ does not increase after performing a local Gaussian
channel on the unmeasured subsystem. Comparing $N_{\mathcal
F}^{\mathcal G}(\rho_{AB})$ in scale with other quantum correlations
such as Gaussian geometric discord for two-mode symmetric squeezed
thermal states reveals that $N_{\mathcal F}^{\mathcal G}$ is  much
better in detecting quantum correlations of Gaussian states.


\end{abstract}

\maketitle

\section{Introduction}

The presence of correlations in bipartite quantum systems  is one of
the main features of quantum mechanics. The most important among
such correlations is surely entanglement \cite{Horodecki}. However,
much attention has been devoted to   studying  and  characterizing
the quantum correlations that go beyond the paradigm of entanglement
recently. Non-entangled quantum correlations are also physical
resources which play   important roles in various quantum
communications and quantum computing tasks.

For the last two decades, various methods have been proposed to
describe quantum correlations, such as quantum discord (QD)
\cite{Ollivier}, geometric quantum discord
\cite{Borivoje,Luo,Miranowicz}, measurement-induced nonlocality
(MIN) \cite{Luo-Fu} and measurement-induced disturbance (MID)
\cite{Luo-S} for discrete-variable systems. For continuous-variable
systems, Giorda, Paris \cite{Giorda} and Adesso, Datta \cite{Adesso}
independently gave the definition of Gaussian QD for two-mode
Gaussian states and discussed its properties.  G. Adesso, D.
Girolami in \cite{Adesso-Girolami} proposed the concept of Gaussian
geometric discord for Gaussian states. Measurement-induced
disturbance of Gaussian states was studied in \cite{Mista}. In
\cite{Ma}, the MIN for Gaussian states was discussed. For other
related results, see \cite{MHQ,Farace, Rigovacca, WHQ, Fu, Datta,
Gharibian} and the references therein. Also, many efforts have been
made to find simpler methods to quantify these correlations.
However, it seems that this is a very difficult task, too. By now,
for example, almost all known quantifications of various
correlations, including entanglement measures, for
continuous-variable systems are difficult to evaluate and can only
be calculated for $(1+1)$-mode Gaussian states or some special
states. Even for finite-dimensional cases, the authors in
\cite{Huang} proved that computing quantum discord is NP-hard.   So
it makes sense and is important to find more helpful quantifications
of quantum correlations.

The purpose of this paper is to propose a correlation $N_{\mathcal
F}^{\mathcal G}$ for bipartite Gaussian systems in terms of local
Gaussian unitary operations based on  square of the fidelity
$\mathcal F$ introduced by Wang, Yu and Yi in \cite{CSYu}. This
correlation $N_{\mathcal F}^{\mathcal G}$ describes the same
correlation as Gaussian geometric discord for Gaussian states but
have some remarkable nice properties that the known quantifications
are not possed: (1) $N_{\mathcal F}^{\mathcal G}$ is a quantum
correlation without ancilla problem; (2) $N_{\mathcal F}^{\mathcal
G}(\rho_{AB})$ can be easily estimated  for any $(n+m)$-mode
Gaussian states and calculated for any $(1+1)$-mode Gaussian states;
(3) $N_{\mathcal F}^{\mathcal G}$ is non-increasing after performing
local Gaussian operations on the unmeasured subsystem. Comparison
$N_{\mathcal F}^{\mathcal G}$ in scale with other quantum
correlations for two-mode symmetric squeezed thermal states reveals
that $N_{\mathcal F}^{\mathcal G}$ is better in detecting the
nonclassicality   in Gaussian states.

\section{Gaussian States and Gaussian unitary operations}

We recall briefly some notions and notations concerning Gaussian
states and Gaussian unitary operations. For arbitrary state $\rho$
in a $n$-mode continuous-variable system with state space $H$, its
characteristic function $\chi_{\rho}$ is defined as
$$\chi_{\rho}(z)={\rm tr}(\rho W(z)),$$ where
$z=(x_{1}, y_{1}, \cdots, x_{n}, y_{n})^{\rm T}\in{\mathbb R}^{2n}$,
$W(z)=\exp(i{R}^{T}z)$ is the Weyl displacement operator,
${R}=(R_1,R_2,\cdots,R_{2n})=(\hat{Q}_1,\hat{P}_1,\cdots,\hat{Q}_n,\hat{P}_n)$.
As usual, $\hat{Q_k}=(\hat{a_k}+\hat{a_k}^\dag)/\sqrt{2}$ and
$\hat{P_k}=-i(\hat{a_k}-\hat{a_k}^\dag)/\sqrt{2}$ ($k=1,2,\cdots,n$)
stand for respectively
 the position and momentum operators, where
 $\hat{a}_k^\dag$ and
$\hat{a}_k$ are the creation and annihilation operators in the $k$th
mode satisfying the Canonical Commutation Relation (CCR)
$$[\hat{a}_k,\hat{a}_l^\dag]=\delta_{kl}I\ {\rm and}
\ [\hat{a}_k^\dag,\hat{a}_l^\dag]=[\hat{a}_k,\hat{a}_l]=0,\ \
k,l=1,2,\cdots,n.$$
 $\rho$ is called a Gaussian state if
$\chi_{\rho}(z)$ is of the form
\begin{eqnarray*}\chi_{\rho}(z)=\exp[-\frac{1}{4}z^{\rm T}\Gamma z+i{\mathbf
d}^{\rm T}z],
\end{eqnarray*}
where  $${\mathbf d}=(\langle\hat R_1 \rangle, \langle\hat R_2
\rangle, \ldots ,\langle\hat R_{2n} \rangle)^{\rm T}=({\rm tr}(\rho
R_1), {\rm tr}(\rho R_2), \ldots, {\rm tr}(\rho R_{2n}))^{\rm
T}\in{\mathbb R}^{2n}$$ is called the mean or the displacement
vector of $\rho$ and $\Gamma=(\gamma_{kl})\in M_{2n}(\mathbb R)$ is
the covariance matrix (CM) of $\rho$ defined by $\gamma_{kl}={\rm
tr}[\rho
(\Delta\hat{R}_k\Delta\hat{R}_l+\Delta\hat{R}_l\Delta\hat{R}_k)]$
with $\Delta\hat{R}_k=\hat{R}_k-\langle\hat{R}_k\rangle$
(\cite{Braunstein}).  Note that  $\Gamma$ is real symmetric and
satisfies the condition $\Gamma +i\Delta\geq 0$, where
$\Delta=\oplus_{k=1}^n\Delta_k$ with $\Delta_k
=\begin{pmatrix}0&1\\-1&0\end{pmatrix}$ for each $k$. Here
$M_d(\mathbb R)$ stands for the algebra of all $d\times d$ matrices
over the real field $\mathbb R$.

Now assume that $\rho_{AB}$ is an  $(n+m)$-mode Gaussian state with
state space $H=H_A\otimes H_B$. Then the CM $\Gamma$ of $\rho_{AB}$
can be written as
\begin{eqnarray}
\Gamma=\left(\begin{array}{cc}A& C\\
C^T & B\end{array}\right),
\end{eqnarray}
where $A \in M_{2n}({\mathbb R})$, $B\in M_{2m}({\mathbb R})$ and
$C\in M_{2n\times 2m}({\mathbb R})$. Particularly, if $n=m=1$, by
means of local  Gaussian  unitary (symplectic at the CM level)
operations, $\Gamma$ has a standard form:
\begin{eqnarray}\Gamma_0=\left(\begin{array}{cc}A_0&C_0\\ C_0^T & B_0\end{array}\right),\end{eqnarray}
where
$A_0=\left(\begin{array}{cc}a&0\\
0 & a\end{array}\right)$, $B_0=\left(\begin{array}{cc}b&0\\
0 & b\end{array}\right)$, $
 C_0=\left(\begin{array}{cc}c&0\\
0 & d\end{array}\right)$, $a,b\geq 1$ and $ab-1\geq c ^2(d^2)$.

For any unitary operator $U$ acting on $H$, the unitary operation
$\rho\mapsto U\rho U^\dag$ is said to be Gaussian if it  sends
Gaussian states into Gaussian states, and such $U$ is called a
Gaussian unitary operator. It is well-known that a unitary operator
$U$ is Gaussian if and only if
\begin{align*}
    U^{\dag}RU = \mathbf SR +\mathbf m
\end{align*}
for some vector $\mathbf m$   in ${\mathbb R}^{2n}$ and some
$\mathbf S \in {\rm Sp}(2n,\mathbb R)$,  the symplectic group of all
$2n\times 2n$ real matrices $\bf S$ that satisfy
$$\mathbf S\in{\rm Sp}(2n,\mathbb R)\Leftrightarrow \mathbf S\Delta \mathbf S^{\rm
T}=\Delta.$$ Thus, every Gaussian unitary operator $U$ is determined
by some affine symplectic map $(\mathbf S, \mathbf m)$ acting on the
phase space, and can be denoted by $U=U_{\mathbf S, \mathbf m}$
(\cite{Wang,Christian}).

We list some simple facts for Gaussian states and Gaussian unitary
operations, and some useful results for matrix theory, which will be
used frequently in the present  paper.

\begin{lemma} {\rm (\cite{Wang})}  For any $(n+m)$-mode Gaussian state
$\rho_{AB}$,  write its CM $\Gamma$ as in Eq.(1). Then the CMs of
the reduced states $\rho_{A}= {\rm tr}_{B}\rho_{AB}$ and $\rho_{B}=
{\rm tr}_{A}\rho_{AB}$ are matrices $A$  and  $B$, respectively.
\end{lemma}

Denote by $S(H)$ the set of all quantum states of the system with
state space $H$.

\begin{lemma} {\rm (\cite{JA})} Assume that $\rho_{AB} \in
S(H_{A}\otimes H_{B})$  is a $(n+m)$-mode Gaussian state. Then
$\rho_{AB}$ is a product state, that is,  $\rho_{AB}
=\sigma_{A}\otimes \sigma_{B}$ for some $\sigma_A\in {\mathcal
S}(H_A)$ and $\sigma_B\in {\mathcal S}(H_B)$, if and only if
$\Gamma=\Gamma_{A }\oplus\Gamma_{B}$, where $\Gamma$, $\Gamma_{A}$
and $\Gamma_{B}$ are the CMs of $\rho_{AB}$, $\sigma_{A}$ and $
\sigma_{B}$, respectively.
\end{lemma}

\begin{lemma} {\rm (\cite{Wang,Christian})} Assume that $\rho$ is
any $n$-mode Gaussian state with CM $\Gamma$ and displacement vector
$\mathbf d$, and assume that $U_{\mathbf S,\mathbf m}$ is a Gaussian
unitary operator. Then the characteristic function of  the Gaussian
state $\sigma=U\rho U^{\dag}$ is of the form
$\exp(-\frac{1}{4}z^{\rm T}\Gamma_{\sigma} z+i\mathbf
d_{\sigma}^{\rm T}z)$, where $\Gamma_{\sigma} = \mathbf S\Gamma
\mathbf S^{\rm T}$ and $\mathbf d_{\sigma} =\mathbf m + \mathbf
S\mathbf d$.
\end{lemma}

\begin{lemma} {\rm (\cite{Holder})}  For any quantum states $\rho$, $\sigma$  and any numbers $a>1$,
we have
$$\tr(\rho\sigma)\leq(\tr \rho^{a})^{\frac{1}{a}}(\tr \sigma^{b})^{\frac{1}{b}},$$
where $b=\frac{a}{a-1}$.
\end{lemma}

\begin{lemma} {\rm (\cite{Horn})}  Let $M =\left(\begin{array}{cc}
A & B \\
C & D
\end{array}
\right)$ be a square matrix.

(1) If $A$ is invertible, then its determinant $\det
\left(\begin{array}{cc}
A & B \\
C & D
\end{array}
\right)=(\det A)( \det(D-CA^{-1}B))$.

(2) If $D$ is invertible, then its determinant $\det
\left(\begin{array}{cc}
A & B \\
C & D
\end{array}
\right)=(\det D )(\det(A-BD^{-1}C))$.
\end{lemma}

\section{Fidelity based nonclassicality of Gaussian states by Gaussian unitary operations}

Fidelity is a measure of closeness between two arbitrary states
$\rho$ and $\sigma$, defined as $F(\rho,\sigma)=(\rm tr
\sqrt{\sqrt{\rho}\sigma\sqrt{\rho}})^{2}$\cite{Jozsa}. This measure
has been explored in various context of quantum information
processing such as cloning \cite{Massar}, teleportation
\cite{GFZhang}, quantum states tomography \cite{Brida}, quantum
chaos \cite{Prosen} and spotlighting phase transition in physical
systems \cite{SJGu}. Though fidelity itself is not a metric, one can
define a metric $D(\rho,\sigma)=g(F(\rho,\sigma)),$ where $g$ is a
monotonically decreasing function of distance measure. A few such
fidelity induced metrics we mentioned here are Bures angle
$A(\rho,\sigma)=\arccos\sqrt{F(\rho,\sigma)}$, Bures metric
$B(\rho,\sigma)=(2-2\sqrt{F(\rho,\sigma)})^{\frac{1}{2}}$ and sine
metric $C(\rho,\sigma)=\sqrt{1-F(\rho,\sigma)}$ \cite{Langford}.

Since the computation of fidelity involves square root of density
matrix, various forms of fidelity have been proposed to simplify the
computation. In \cite{CSYu}, the authors proposed another form
$\mathcal F$ of   fidelity as
\begin{eqnarray}
 \mathcal F(\rho,\sigma)=\frac{|\rm tr\rho\sigma|}{\sqrt{\rm tr
\rho^{2} \rm tr \sigma^{2}}},
\end{eqnarray} In \cite{RMRS}, to capture global nonlocal effect of a
quantum state of discrete system due to locally invariant projective
measurements, the authors use the fidelity in Eq.(3) to define a
metric $C(\rho,\sigma)=\sqrt{1-\mathcal F^2(\rho,\sigma)}$ for any
states $\rho$ and $\sigma$. Furthermore, for any finite-dimensional
bipartite quantum state $\rho_{AB}$,  a new kind of MIN in terms of
this metric was defined as
$$N_{\mathcal
F}(\rho_{AB})=\max\limits_{\Pi^{A}}C^{2}(\rho_{AB},\Pi^{A}(\rho_{AB})),$$
where the maximum is taken over all von Neumann measurements
performing on subsystem A that are invariant at $\rho_A={\rm
tr}_B(\rho_{AB})$, the reduced state of $\rho_{AB}$.   They
presented an analytic expression of this version of MIN for pure
bipartite states and $2\times n$ dimensional mixed states.

In the present  paper, motivated by the work of \cite{RMRS}, we
propose a quantum nonclassicality $N_{\mathcal F}^{\mathcal G}$ for
continuous-variable systems by local Gaussian unitary operations for
$(n+m)$-mode states using the same metric based on the fidelity
Eq.(3).

\begin{definition} For any $(n+m)$-mode state $\rho_{AB}\in \mathcal
S(H_{A}\otimes H_{B})$, the quantity $N_{\mathcal F}^{\mathcal
G}(\rho_{AB})$ is defined by
\begin{eqnarray}
 N_{\mathcal F}^{\mathcal G}(\rho_{AB}) = \sup_{U} C^{2}(\rho_{AB},
(U\otimes
\textit{I})\rho_{AB}(U^{\dag}\otimes\textit{I}))=\sup_{U}\{1-\frac{({\rm
tr}\rho_{AB}(U\otimes \textit{I})\rho_{AB}(U^{\dag}\otimes
\textit{I}))^{2}} {{\rm tr}(\rho_{AB}^{2}){\rm tr}((U\otimes
\textit{I})\rho_{AB}(U^{\dag}\otimes \textit{I}))^{2}} \},
\end{eqnarray}
where the supremum is taken over all Gaussian unitary operators $U$
on $H_A$ satisfying $U\rho_{A}U^{\dag}=\rho_{A}$.
\end{definition}

\begin{remark}
For any Gaussian state $\rho_{AB}$, there are many  nontrivial
Gaussian unitary operators $U$ (other than the identity $I$)
satisfying $U\rho_{A}U^{\dag}=\rho_{A}$ \cite {WHQ}, and hence
Definition 1 makes sense.
 Different from \cite{WHQ}, in which a quantum nonclassicality $\mathcal N$ is proposed by Gaussian unitary operations based
 on the Hilbert-Schmidt norm, the quantity $N_{\mathcal
F}^{\mathcal G}(\rho_{AB})$ measures the global nonlocal effect of a
quantum state due to locally invariant Gaussian unitary operations
by the metric $C^2(\rho,\sigma)={1-\mathcal F^2(\rho,\sigma)}$ with
the fidelity $\mathcal F$ as in Eq.(3).
\end{remark}

Recall that, the  MIN \cite {Luo-Fu} is defined as the square of
Hilbert-Schmidt norm $\|\cdot\|_2$ ($\|A\|_2=\sqrt{{\rm tr}(A^\dag
A)}\ $) of difference of pre- and post-measurement states. i.e.,
$$N(\rho_{AB})=\max_{\Pi^{A}}\|\rho_{AB}-(\Pi^{A}\otimes
I)\rho_{AB}(\Pi^{A}\otimes I)^{\dagger}\|_2^{2},$$ where the maximum
is taken over all von Neumann measurements which maintain the
reduced state $\rho_A$ invariant corresponding to part A. In \cite
{WHQ}, a kind of quantum correlation $\mathcal N$ for $(n+m)$-mode
continuous-variable systems is defined as the square of
Hilbert-Schmidt norm of difference of pre- and post-transform states
$$\mathcal N(\rho_{AB})=\frac{1}{2}\sup_{U}\|\rho_{AB}-(U\otimes I)\rho_{AB}(U\otimes
I)^{\dagger}\|^{2}_2,$$ where the supremum is taken over all unitary
operators which maintain  $\rho_A$  invariant corresponding to party
A. There are other quantum correlations defined by Hilbert-Schmidt
norm, for example, the  Gaussian geometric discord and the quantum
correlation  proposed respectively in \cite{Adesso-Girolami,MHQ}.
These kinds of quantity defined by Hilbert-Schmidt norm mentioned
above may change rather wildly through some trivial and uncorrelated
actions on the unoperated party B. For example, if we append an
uncorrelated ancilla C, and regarding the state
$\rho_{ABC}=\rho_{AB}\otimes \rho_{C}$ as a bipartite state with the
partition A:BC. After some straight-forward calculations, one   gets
$$ \mathcal N(\rho_{ABC})=\mathcal N(\rho_{AB}){\rm
tr}\rho_{C}^{2},$$ which means that the quantity $\mathcal N$
differs arbitrarily due to local ancilla C as long as $\rho_{C}$ is
mixed. While  this problem can be avoided if one employs ${\mathcal
N}_{\mathcal F}^{\mathcal G}$ as in Definition 1 since
\begin{align*}
&\mathcal F(\rho_{ABC},(U\otimes I\otimes I)\rho_{ABC}) =\mathcal
F(\rho_{AB}\otimes \rho_{C},(U\otimes I)\rho_{AB}\otimes
I\rho_{C})\\=& \mathcal F(\rho_{AB},(U\otimes I)\rho_{AB})\cdot
\mathcal F(\rho_{C},\rho_{C})=\mathcal F(\rho_{AB},(U\otimes
I)\rho_{AB}),
\end{align*}
according to the multiplicativity of the fidelity \cite{CSYu}. Thus,
we reach  the following conclusion.

\begin{theorem}
$N_{\mathcal F}^{\mathcal G}$ is a quantum nonclassicality without
ancilla problem.
\end{theorem}

We explore further the properties of ${  N}_{\mathcal F}^{\mathcal
G}$ below. Denote by ${\mathcal B}(H)$ the algebra of all bounded
linear operators acting on $H$.

\begin{theorem}
$ N_{\mathcal F}^{\mathcal G}$ is locally Gaussian unitary
invariant, that is, for any $(n+m)$-mode Gaussian state
$\rho_{AB}\in \mathcal S(H_{A}\otimes H_{B})$ and any Gaussian
unitary operators $W\in{\mathcal B}(H_A)$ and $V\in{\mathcal
B}(H_{B})$, we have $N_{\mathcal F}^{\mathcal G}((W\otimes
V)\rho_{AB}(W^{\dag}\otimes V^{\dag})) = N_{\mathcal F}^{\mathcal
G}(\rho_{AB}) $.
\end{theorem}

{\bf Proof.} Assume that $\rho_{AB}\in \mathcal S(H_{A}\otimes
H_{B})$ is an $(n+m)$-mode Gaussian state. For given Gaussian
unitary operators $W\in{\mathcal B}(H_A)$ and $V\in{\mathcal
B}(H_{B})$, let $\sigma_{AB} =(W\otimes V)\rho_{AB} (W^{\dag}\otimes
V^{\dag})$. Denote ${\mathcal {U}}^{\mathcal G}(H_A)$ the set of all
Gaussian unitary operators acting on $H_A$. Since
\begin{align*} N_{\mathcal F}^{\mathcal G}(\rho_{AB}) = &
\sup_{U\in{\mathcal{U}^{\mathcal G}(H_A)},\ U\rho_AU^\dag=\rho_A}
C^{2}(\rho_{AB}, (U\otimes
\textit{I})\rho_{AB}(U^{\dag}\otimes\textit{I}))\\
=& \sup_{U\in{\mathcal{U}^{\mathcal G}(H_A)},\
U\rho_AU^\dag=\rho_A}\{1-\mathcal F^2(\rho_{AB},(U\otimes
I)\rho_{AB}(U^{\dag}\otimes I))\}\\ = &
1-\inf_{U\in{\mathcal{U}^{\mathcal G}(H_A)},\
U\rho_AU^\dag=\rho_A}\mathcal F^2(\rho_{AB},(U\otimes
I)\rho_{AB}(U^{\dag}\otimes I)) ,
\end{align*} to demonstrate that $N_{\mathcal F}^\mathcal G$ is locally Gaussian unitary invariant,
it is sufficient to prove
\begin{align} & \inf_{U\in{\mathcal{U}^{\mathcal G}(H_A)},\ U\rho_AU^\dag=\rho_A}\mathcal F(\rho_{AB},(U\otimes
I)\rho_{AB}(U^{\dag}\otimes I))\\\nonumber= & \inf_{ U^\prime
\in{\mathcal{U}^{\mathcal G}(H_A)},\ U^\prime\sigma_A
U^{\prime\dag}=\sigma_A }\mathcal F(\sigma_{AB},(U^{\prime}\otimes
I)\sigma_{AB}(U^{\prime\dag}\otimes I)),
\end{align} where $\sigma_{AB}=(W\otimes
V)\rho_{AB}(W^{\dag}\otimes V^{\dag})$,
 $W$ and $V$ are given Gaussian unitary operators acting on
Hilbert spaces $H_{A}$ and $H_{B}$, respectively.

Note that  $\sigma_{A} = W\rho_{A}W^{\dag}$.
 For any Gaussian unitary operator $U\in{\mathcal
B}(H_{A})$ satisfying $U\rho_{A} U^{\dag}=\rho_{A}$, let $U^{\prime}
= WUW^{\dag}$.   Then $U^\prime$ is  a Gaussian unitary operator
satisfing $U^{\prime}\sigma_{A}U^{\prime\dag} =WUW^{\dag}W\rho_{A}
W^{\dag} WU^{\dag}W^\dag =\sigma_{A}$. Conversely, if $U^\prime
\sigma_{A}U^{\prime\dag}=\sigma_{A}$, $U=W^\dag U^\prime W$ will
satisfy $U\rho_AU^\dag=\rho_A$. By Eq.(3), we have
\begin{align*}
&\mathcal F^2(\rho_{AB},(U\otimes I)\rho_{AB}(U^{\dag}\otimes I))=\frac{(\tr \rho_{AB}(U\otimes I)\rho_{AB}(U^{\dag}\otimes I))^{2}}{\tr \rho_{AB}^{2}\tr((U\otimes I)\rho_{AB}(U^{\dag}\otimes I))^{2}} \\
=& \frac{(\tr (W^\dag\otimes V^\dag)\sigma_{AB} (W \otimes V
)(U\otimes I)(W^\dag\otimes V^\dag)\sigma_{AB} (W \otimes V
)(U^{\dag}\otimes I))^{2}}
{\tr ((W^\dag\otimes V^\dag)\sigma_{AB} (W \otimes V ))^{2}\tr((U\otimes I)(W^\dag\otimes V^\dag)\sigma_{AB} (W \otimes V )(U^{\dag}\otimes I))^{2}} \\
=&\frac{(\tr \sigma_{AB} (U^{\prime}\otimes I)\sigma_{AB}
(U^{\prime\dag}\otimes I))^{2}} {\tr
\sigma_{AB}^{2}\tr((U^{\prime}\otimes
I)\sigma_{AB}(U^{\prime\dag}\otimes I))^{2}} =\mathcal
F^2(\sigma_{AB},(U^{\prime}\otimes
I)\sigma_{AB}(U^{\prime\dag}\otimes I)).
\end{align*}
Therefore, Eq.(5) holds, as desired. \hfill$\Box$

Notice that, for any $(n+m)$-mode product quantum state $\rho_{AB}$,
one must have $N_{\mathcal F}^{\mathcal G}(\rho_{AB})=0$ by the
definition. But for Gaussian states, the converse is also true.
Hence, when restricted to Gaussian states, the correlation
$N_{\mathcal F}^{\mathcal G}$ describes the same nonclassicality as
that described by Gaussian QD (two-mode) \cite{Giorda, Adesso},
Gaussian geometric discord \cite{Adesso-Girolami}, the correlations
$Q$, $Q_{\mathcal P}$ discussed in \cite{MHQ} and the correlation
${\mathcal N}$ discussed in \cite{WHQ}.

\begin{theorem}
For any $(n+m)$-mode Gaussian state $\rho_{AB}\in {\mathcal
S}(H_A\otimes H_B)$, $N_{\mathcal F}^{\mathcal G}(\rho_{AB}) = 0$ if
and only if $\rho_{AB}$ is a product state.
\end{theorem}

{\bf Proof.} By Definition 1, the ``if'' part is apparent. Let us
check the ``only if'' part. Since the mean of any Gaussian state can
be transformed to zero under some local Gaussian unitary operation,
by Theorem 2, it is sufficient to consider the Gaussian states whose
mean are zero.

 Assume that $\rho_{AB}$ is an $(n+m)$-mode Gaussian
state with CM $\Gamma =\left(\begin{array}{cc}
A & C \\
C^{T} & B
\end{array}
\right)$ as in Eq.(1) and zero mean such that $N_{\mathcal
F}^{\mathcal G}(\rho_{AB})=0$. By Lemma 1, the CM of $\rho_A$ is
$A$. According to the Williamson Theorem, there exists a symplectic
matrix $\mathbf S_{0}$ such that $\mathbf S_{0}A\mathbf S_{0}^{\rm
T} = \oplus_{i=1}^n v_{i}I$ and $U_{0}\rho_{A}
U_{0}^{\dag}=\otimes_{i=1}^n\rho_{i}$, where $U_0=U_{{\bf S}_0,{\bf
0}}$ and  $\rho_{i}$s are some thermal states. Write
$\sigma_{AB}=(U_{0}\otimes I)\rho_{AB}(U_{0}^{\dag}\otimes I)$. It
follows from Theorem 2 that $N_{\mathcal F}^{\mathcal
G}(\sigma_{AB})=N_{\mathcal F}^{\mathcal G}(\rho_{AB})=0$.
Obviously, $\sigma_{AB}$ has the CM
$$\Gamma^{\prime} = \left(\begin{array}{cc}
\oplus_{i}^nv_{i}I & C^{\prime} \\
C^{\prime \rm T} & B^{\prime}
\end{array}
\right)$$ and the mean 0.

By Lemma 3 and \cite{WHQ}, for any Gaussian unitary operator
$U_{\mathbf S,\mathbf m}\in{\mathcal B}(H_A)$ so that $\mathbf m=0$
and $\mathbf S=\oplus_{i=1}^n \mathbf S_{\theta_{i}}$ with $$
\mathbf S_{\theta_{i}} = \left(
\begin{array}{cc}
\cos\theta_{i} & \sin\theta_{i} \\
-\sin\theta_{i} & \cos\theta_{i} \\
\end{array}
\right)$$ for some $\theta_{i}\in [0, \frac{\pi}{2}]$, we have
$U_{\mathbf S,\mathbf m}\sigma_{A}U_{\mathbf S,\mathbf m}^{\dag} =
\sigma_{A}={\rm tr}_B(\sigma_{AB})$. Then, by the definition Eq.(4),
$N_{\mathcal F}^{\mathcal G}(\sigma_{AB})=0$ entails
$$(\rm tr\sigma_{AB}(U_{\mathbf S,\mathbf m}\otimes
I)\sigma_{AB}(U_{\mathbf S,\mathbf m}^\dag\otimes I))^{2}=\rm
tr\sigma_{AB}^{2}{\rm tr}((U_{\mathbf S,\mathbf m}\otimes
I)\sigma_{AB}(U_{\mathbf S,\mathbf m}^\dag\otimes I))^{2}.$$ Since
the Holder's inequality (Lemma 4) asserts that ${\rm
tr}(\rho\sigma)^2\leq{\rm tr}\rho^2{\rm tr}\sigma^2$  and clearly,
the equality holds if and only if $\sigma=\rho$, we must have
$$\sigma_{AB}=(U_{\mathbf S,\mathbf m}\otimes
I)\sigma_{AB}(U_{\mathbf S,\mathbf m}^\dag\otimes I).$$ Hence
$\sigma_{AB}$ and $(U_{\mathbf S,\mathbf m}\otimes
I)\sigma_{AB}(U_{\mathbf S,\mathbf m}^\dag\otimes I)$   have the
same CMs, that is,
\begin{align*}
\left(
\begin{array}{cc}
\oplus_{i=1}^nv_{i}I & C^{\prime} \\
C^{\prime \rm T} & B^{\prime}
\end{array}
\right) = \left(
\begin{array}{cc}
 \oplus_{i=1}^nv_{i}I & {\bf S}C^{\prime} \\
 C^{\prime \rm T}{\bf S}^{\rm T} & B^{\prime} \\
\end{array}
\right).
\end{align*}
If  we take  $\theta_{i}\in(0, \frac{\pi}{2})$ for each $i$, then $I
- {\bf S}$ is an invertible matrix, which forces $C^{\prime} = 0.$
So $\sigma_{AB}$ is a product  state by Lemma 2. It follows that
$\rho_{AB} = (U_0^{\dag} \otimes I )\sigma_{AB}(U_0 \otimes I)$ is
also a product state. \hfill$\Box$

  In the rest of this paper, we mainly consider the case when the
states $\rho_{AB}$ are Gaussian.

A remarkable virtue  of $N_{\mathcal F}^{\mathcal G}$ is that it can
be evaluated easily. For any two-mode Gaussian state $\rho_{AB}$, we
can give an analytic computation formula.

\begin{theorem}
For any $(1+1)$-mode  Gaussian state $\rho_{AB}$ whose CM has the
standard form  $\Gamma_0=\left(\begin{array}{cc}
A_{0} & C_{0} \\
C_{0}^{\rm T} & B_{0}
\end{array}
\right)=\left(\begin{array}{cccc}a & 0 & c & 0\\
0 & a & 0 & d\\
c & 0 & b & 0\\
0 & d & 0 & b\end{array}\right)$ , we have
\begin{align*}
{  N}_{\mathcal F}^{\mathcal G}(\rho_{AB}) =
1-\frac{(ab-c^{2})(ab-d^{2})}{(ab-c^{2}/2)(ab-d^{2}/2)}.
\end{align*}
Particularly, the value of ${ N}_{\mathcal F}^{\mathcal
G}(\rho_{AB})$ is independent of the mean of the state $\rho_{AB}$.
\end{theorem}

{\bf Proof.}  For any $(1+1)$-mode Gaussian state $\rho_{AB}$ with
CM $\Gamma^{\prime}$ and mean $(\mathbf d_{A}^{\prime}, \mathbf
d_{B}^{\prime})$, we can always find two Gaussian operators $U$ and
$V$ so that the CM $\Gamma_0$ of $\sigma_{AB}=(U\otimes
V)\rho_{AB}(U^{\dag}\otimes V^{\dag})$  is of the standard form
$$\Gamma_0=\left(\begin{array}{cc}
A_{0} & C_{0} \\
C_{0}^{\rm T} & B_{0}
\end{array}
\right)=\left(\begin{array}{cccc}a & 0 & c & 0\\
0 & a & 0 & d\\
c & 0 & b & 0\\
0 & d & 0 & b\end{array}\right).$$ Denote  the mean of $\sigma_{AB}$
by $(\mathbf d_{A}, \mathbf d_{B})$. Since $N_{\mathcal F}^{\mathcal
G}$ is locally Gaussian unitary invariant, one has $N_{\mathcal
F}^{\mathcal G}(\rho_{AB})=N_{\mathcal F}^{\mathcal
G}(\sigma_{AB})$. Hence, we may assume that the CM of $\rho_{AB}$ is
$\Gamma_0$ and the mean of $\rho_{AB}$ is  $(\mathbf d_{A}, \mathbf
d_{B})$. For any Gaussian unitary operator $U_{\mathbf S,\mathbf m}$
such that $U_{\mathbf S,\mathbf m}\rho_{A} U_{\mathbf S,\mathbf
m}^{\dag}=\rho_{A}$, we see that $\mathbf S$ and $\mathbf m$ meet
the conditions $\mathbf SA_0\mathbf S^{\rm T} =A_0$ and $ \mathbf
S\mathbf d_{A} + \mathbf m=\mathbf d_{A}$. As $A_0=aI_2$, we have
$\mathbf S\mathbf S^{\rm T}=I_2$.  It follows from $\mathbf S\Delta
\mathbf S^{\rm T} = \Delta $  that there exists some $\theta\in [0,
\frac{\pi}{2}]$ such that ${\bf S} = {\bf S}_{\theta} = \left(
\begin{array}{cc}
\cos\theta & \sin\theta \\
-\sin\theta & \cos\theta
\end{array}
\right)$. So the CM of Gaussian state $(U_{\mathbf S,\mathbf
m}\otimes I)\rho_{AB}(U_{\mathbf S,\mathbf m}^{\dag}\otimes I)$ is
\begin{align*}
\Gamma_{\theta}= \left(
\begin{array}{cccc}
a & 0 & c\cos\theta & d\sin\theta \\
0 & a & -c\sin\theta & d\cos\theta \\
c\cos\theta & -c\sin\theta & b & 0 \\
d\sin\theta & d\cos\theta & 0 & b
\end{array}
\right),
\end{align*}
and the mean of $(U_{\mathbf S,\mathbf m}\otimes
I)\rho_{AB}(U_{\mathbf S,\mathbf m}^{\dag}\otimes I)$ is $$({\bf
S}\oplus I)(\mathbf d_{A}\oplus \mathbf d_{B})+\mathbf m\oplus
0=({\bf S} \mathbf d_{A}+\mathbf m) \oplus \mathbf d_{B}=\mathbf
d_{A}\oplus \mathbf d_{B}=(\mathbf d_{A}, \mathbf d_{B})$$ as
$\mathbf S\mathbf d_{A} + \mathbf m=\mathbf d_{A}$. Conversely, for
any $\mathbf S_\theta$, taking $\mathbf m=\mathbf d_A-\mathbf
S_\theta \mathbf d_A$, we have $U_{\mathbf S_\theta,\mathbf m}$
satisfies the condition $U_{\mathbf S_\theta,\mathbf m}\rho_{A}
U_{\mathbf S_\theta,\mathbf m}^{\dag}=\rho_{A}$.

Also, notice that, for any $n$-mode Gaussian states $\rho,\sigma$
with CMs $V_{\rho}, V_{\sigma}$ and means $\mathbf d_{\rho}, \mathbf
d_{\sigma}$, respectively, it is shown in \cite{Marian} that
\begin{widetext}\begin{align}\textrm{Tr}(\rho\sigma) =
\frac{1}{\sqrt{\det[(V_{\rho}+V_{\sigma})/2]}}\exp[-\frac{1}{2}\delta\langle
\mathbf d\rangle^{T}\det[(V_{\rho}+V_{\sigma})/2]^{-1}\delta\langle
\mathbf d\rangle],\end{align}\end{widetext}  where $ \delta\langle
\mathbf d\rangle =\mathbf d_{\rho} - \mathbf d_{\sigma}$.

Hence, by Eq.(4) and Eq.(6) as well as the fact that
$\det\Gamma_\theta=\det\Gamma_0=(ab-c^{2})(ab-d^{2})$, one obtains
\begin{align*}
N_{\mathcal F}^{\mathcal G}(\rho_{AB})
=&\sup_{U\in{\mathcal{U}^{\mathcal G}(H_A)},\
U\rho_AU^\dag=\rho_A}C^{2}(\rho_{AB}, (U\otimes \textit{I})\rho_{AB}(U^{\dag}\otimes \textit{I}))\\
= & \sup_{U\in{\mathcal{U}^{\mathcal G}(H_A)},\
U\rho_AU^\dag=\rho_A} \{1-\frac{(\textrm{tr}\rho_{AB}(U\otimes
\textit{I})\rho_{AB}(U^{\dag}\otimes \textit{I}))^{2}}
{\textrm{tr}(\rho_{AB}^{2})\textrm{tr}((U\otimes
\textit{I})\rho_{AB}(U^{\dag}\otimes \textit{I}))^{2}} \} \\
= &\sup_{\theta\in [0, \frac{\pi}{2}]}\{1-\frac{\sqrt{\det\Gamma_0\det\Gamma_\theta}}{{\det((\Gamma_0 +\Gamma_\theta)/2)}} \} \\
= & \max_{\theta\in [0, \frac{\pi}{2}]}\{1-
\frac{(ab-c^{2})(ab-d^{2})}{{[ab-c^{2}(1+\cos\theta)/2][ab-d^{2}(1+\cos\theta)/2]}}\} \\
= & 1-\frac{(ab-c^{2})(ab-d^{2})}{{(ab-c^{2}/2)(ab-d^{2}/2)}},
\end{align*}
and, this quantity is independent of the mean of $\rho_{AB}$,
completing the proof. \hfill$\Box$

Next, we are going to give an estimate of $N_{\mathcal F}^{\mathcal
G}$ for any $(n+m)$-mode Gaussian state $\rho_{AB}$.

\begin{theorem}
For any  $(n+m)$-mode Gaussian state $\rho_{AB}$ with CM $\Gamma
=\left(\begin{array}{cc}
A & C \\
C^{\rm T} & B
\end{array}
\right)$, $ N_{\mathcal F}^{\mathcal G}(\rho_{AB})$ is independent
of the mean of $\rho_{AB}$ and
\begin{align*}
0 \leq N_{\mathcal F}^{\mathcal G}(\rho_{AB}) \leq 1- \frac{\det
(B-C^{\rm T}A^{-1}C)}{{\det B}}< 1.
\end{align*}
Furthermore, the upper bound $1$ is tight.
\end{theorem}

{\bf Proof.} Let $\rho_{AB}$ be any  $(n+m)$-mode Gaussian state
with CM $\Gamma =\left(\begin{array}{cc}
A & C \\
C^{\rm T} & B
\end{array}
\right)$ and mean $\mathbf d=(\mathbf d_{A},\mathbf d_{B})$. Note
that, by Lemma 1, the CM of $\rho_A$ is $A$. Write
$\sigma_{AB}=(U_{\mathbf S,\mathbf m}\otimes I)\rho_{AB}(U_{\mathbf
S,\mathbf m}^{\dag}\otimes I)$, where $U_{\mathbf S,\mathbf m}$ is
any Gaussian unitary operator of the subsystem $A$. Clearly,
 $U_{\mathbf S,\mathbf m}\rho_{A} U_{\mathbf S,\mathbf
m}^{\dag} = \rho_{A}$   if and only if the symplectic matrix
$\mathbf S$ satisfies $\mathbf S A \mathbf S^{\rm T} = A$ and the
vector $\mathbf m=\mathbf d_{A}- \mathbf S\mathbf d_{A}  $. In this
case $\sigma_{AB}$ has the CM
$$\Gamma_{\mathbf S}= \left(
\begin{array}{cc}
A & {\bf S}C \\
C^{\rm T}{\bf S}^{\rm T} & B
\end{array}
\right)$$ and the mean $\mathbf d_{\mathbf S}=({\bf S}\oplus
I)(\mathbf d_{A}\oplus \mathbf d_{B})+\mathbf m\oplus 0=({\bf S}
\mathbf d_{A}+\mathbf m) \oplus \mathbf d_{B}=\mathbf d_{A}\oplus
\mathbf d_{B}=(\mathbf d_{A}, \mathbf d_{B})=\mathbf d$. Denote by
$\mathcal S (2n)={\rm Sp}(2n, \mathbb R),$ the set of all $2n\times
2n$ symplectic matrices. Then, by Eq.(6),
\begin{align*}
N_{\mathcal F}^{\mathcal G}(\rho_{AB}) = &
\sup_{U\in{\mathcal{U}^{\mathcal G}(H_A)},\
U\rho_AU^\dag=\rho_A}C^{2}(\rho_{AB},
(U\otimes \textit{I})\rho_{AB}(U^{\dag}\otimes \textit{I})) \\
=&\sup_{U\in{\mathcal{U}^{\mathcal G}(H_A)},\
U\rho_AU^\dag=\rho_A}\{1-\frac{(\textrm{tr}\rho_{AB}(U\otimes
\textit{I})\rho_{AB}(U^{\dag}\otimes \textit{I}))^{2}}
{\textrm{tr}(\rho_{AB}^{2})\textrm{tr}((U\otimes \textit{I})\rho_{AB}(U^{\dag}\otimes \textit{I}))^{2}} \}\\
 = & \sup_{\mathbf S\in{\mathcal S}(2n),\ \mathbf S A
\mathbf S^{\rm T} = A}\{1-\frac{\frac{1}{({\det(\Gamma +\Gamma_{\bf S})/2)}}}{\frac{1}{\sqrt{\det\Gamma}}\frac{1}{\sqrt{\det\Gamma_{\bf S}}}} \}\\
 =& \sup_{\mathbf S\in{\mathcal S}(2n),\ \mathbf S A
\mathbf S^{\rm T} = A}\{1-\frac{\sqrt{\det\Gamma\det\Gamma_{\bf
S}}}{{\det((\Gamma +\Gamma_{\bf S})/2)}} \}.
\end{align*}
That is,
\begin{align}
N_{\mathcal F}^{\mathcal G}(\rho_{AB}) =\sup_{\mathbf S\in{\mathcal
S}(2n),\ \mathbf S A \mathbf S^{\rm T} =
A}\{1-\frac{\sqrt{\det\Gamma\det\Gamma_{\bf S}}}{{\det((\Gamma
+\Gamma_{\bf S})/2)}} \}.
\end{align}
Obviously, $N_{\mathcal F}^{\mathcal G}(\rho_{AB})$ is independent
of the mean $\mathbf d$.

It is easy to verify that $\det \Gamma=\det \Gamma_{\bf S}$. Since
$\Gamma= \left(
\begin{array}{cc}
A & C \\
C^{\rm T} & B
\end{array} \right)>0$, by Lemma
4, we have $$0<\det\Gamma=\det A\det(B-C^{\rm
T}A^{-1}C)=\det\Gamma_{\bf S},$$ which implies that $\det(B-C^{\rm
T}A^{-1}C)>0$. In addition, as $\frac{\Gamma + \Gamma_{\bf S}}{2}=
\left(\begin{array}{cc}
A & \frac{C +{\bf S}C}{2} \\
\frac{C^{ T}+C^{ T}{\bf S}^{T}}{2}& B
\end{array}
\right)$ and  $A$, $B$ are positive-definite, by Fischer's
inequality (\cite[pp.506]{Horn}), we have $\det\frac{\Gamma +
\Gamma_{\bf S}}{2}\leq \det A\det B$. Hence, by Eq.(7), we get
\begin{align*}
0\leq N_{\mathcal F}^{\mathcal G}(\rho_{AB}) \leq & 1- \frac{{\det
A\det(B-C^{\rm T}A^{-1}C)}}{{\det A\det B }}=1- \frac{\det (B-C^{\rm
T}A^{-1}C)}{{\det B}}< 1.\end{align*}

We claim that the upper bound $1$ is tight, that is, we have
\begin{align} \sup_{\rho_{AB}} N_{\mathcal F}^{\mathcal G}(\rho_{AB})=1.\end{align} To see this, consider a two-mode squeezed vacuum state
$\rho(r) = S(r)|00\rangle\langle 00|S^{\dag}(r)$, where
$S(r)=\exp(-r\hat{a}_{1}\hat{a}_{2}+r\hat{a}_{1}^{\dag}\hat{a}_{2}^{\dag})$
is a two-mode squeezing operator with squeezed number $r\geq 0$ and
$|00\rangle$ is the vacuum state (\cite{Milburn}). The CM of
$\rho(r)$ is $\frac{1}{2}\left(\begin{array}{cc}
A_r&  C_r \\
C_r^{ T} & B_r
\end{array}
\right),$ where $$A_r=B_r=\left(\begin{array}{cc}
\exp(-2r)+\exp(2r) & 0\\
0 & \exp(-2r)+\exp(2r)
\end{array}
\right)$$   and  $$C_r=C_r^{ T}=\left(\begin{array}{cc}
 -\exp(-2r)+\exp(2r) & 0 \\
0 & \exp(-2r)-\exp(2r)
\end{array}
\right).$$ By Theorem 4, it is easily checked that
$$N_{\mathcal F}^{\mathcal G}(\rho(r))=1-\frac{16}{(\frac{\exp(-4r)+\exp(4r)}{2}+3)^{2}}.$$  Clearly,
$N_{\mathcal F}^{\mathcal G}(\rho(r)) \rightarrow 1$ as $r
\rightarrow\infty$. So $ \sup_r N_{\mathcal F}^{\mathcal
G}(\rho(r))=1$ and Eq.(8) is true. \hfill$\Box$

Suppose that $\rho_{AB}$ is an $(n+m)$-mode Gaussian state with CM
$\Gamma =\left(\begin{array}{cc}
A & C \\
C^{\rm T} & B
\end{array}
\right)$ as in Eq.(1). One can always perform a local Gaussian
unitary operation on the state $\rho_{AB}$, say
$\sigma_{AB}=(U_{S_A}\otimes V_{S_B})\rho_{AB}(U_{S_A}^{\dag}\otimes
V_{S_B}^{\dag})$, such that the corresponding CM of $\sigma_{AB}$ is
of the form $\Gamma^{\prime} = \left(\begin{array}{cc}
\oplus_{i}^nv_{i}I_2 & C^{\prime} \\
C^{\prime \rm T} & \oplus_{i}^ms_{i}I_2
\end{array}
\right)$, where $v_{i}$s and $s_{i}$s are the symplectic roots of
 $\rho_A$ and $\rho_B$ respectively, $C^\prime=S_ACS_B^{\rm T}$. By Theorem 2,
$N_{\mathcal F}^{\mathcal G}(\sigma_{AB})=N_{\mathcal F}^{\mathcal
G}(\rho_{AB})$. This gives an  estimation of $N_{\mathcal
F}^{\mathcal G}(\rho_{AB})$ for $(n+m)$-mode Gaussian state
$\rho_{AB}$ in terms of symplectic roots of the CMs of the reduced
states $\rho_A$ and $\rho_B$:
\begin{align*}
0 \leq N_{\mathcal F}^{\mathcal G}(\rho_{AB}) \leq 1- \frac{\det
(\oplus_{i}^m s_{i}   I_2-S_BC^{\rm T}S_A^{\rm
T}(\oplus_{i}^n1/v_{i}   I_2)S_ACS_B^{\rm
T})}{{\prod_{i=1}^{m}s_{i}^{2}}}< 1.
\end{align*}

\section{Nonlocality connected to Gaussian channels}
In this section we intend to investigate the fidelity based
nonlocality connected to a Gaussian quantum channel. Here we mainly
consider the $(1+1)$-mode Gaussian states whose CM are of the
standard form.

Since a Gaussian state $\rho$ is described by its CM $\Gamma$ and
displacement vector $\mathbf d$, we can denote it  as
$\rho=\rho(\Gamma,\mathbf d)$. Recall that a Gaussian channel is a
quantum channel that transforms Gaussian states into Gaussian
states. Assume that $\Phi$ is a Gaussian channel of $n$-mode
Gaussian systems. Then, there exist real matrices $M, K\in
M_{2n}(\mathbb R)$ satisfying $M=M^T\geq 0$ and det$ M\geq ({\rm det
K}-1)^2$, and a vector  $\overline{\mathbf d}\in {\mathbb R}^{2n}$,
such that, for any $n$-mode Gaussian state $\rho=\rho(\Gamma,\mathbf
d)$, we have $\Phi(\rho(\Gamma,\mathbf
d))=\rho(\Gamma^{\prime},\mathbf d^{\prime})$ with
$$\mathbf d^{\prime}=K\mathbf d+\overline{\mathbf d}\ \ {\rm and}\ \ \Gamma^{\prime}=K\Gamma K^{T}+M.$$ So we can parameterize the
Gaussian channel $\Phi$ as $\Phi=\Phi(K, M, \overline{\mathbf d})$.

\begin{theorem}
Consider the $(1+1)$-mode continuous-variable system AB. Let
$\Phi=\Phi(K, M, \overline{\mathbf d})$ be a Gaussian channel  performed on the subsystem B with $K=\left(\begin{array}{cc}k_{11}&k_{12}\\
k_{21} & k_{22}\end{array}\right)$ and $M=\left(\begin{array}{cc}m_{11} &m_{12}\\
m_{12} & m_{22}\end{array}\right)$. Assume that
$\rho_{AB}\in{\mathcal S}(H_A\otimes H_B)$ is any $(1+1)$-mode
Gaussian state with CM  $\Gamma_0=\left(\begin{array}{cccc}a & 0 & c & 0\\
0 & a & 0 & d\\
c & 0 & b & 0\\
0 & d & 0 & b\end{array}\right)$. Then
\begin{widetext}\begin{align*}
N_{\mathcal F}^{\mathcal G}((I\otimes
\Phi)\rho_{AB})=1-\frac{(ab-c^{2})(ab-d^{2})n_{1}+a(ab-c^{2})n_{2}+a(ab-d^{2})n_{3}+a^{2}n_{4}}
{(ab-c^{2}/2)(ab-d^{2}/2)n_{1}+a(ab-c^{2}/2)n_{2}+a(ab-d^{2}/2)n_{3}+a^{2}n_{4}},
\end{align*}\end{widetext}
where $n_{0}=(1+\cos\theta)/2$,
$n_{1}=k^{2}_{11}k^{2}_{22}+k^{2}_{12}k^{2}_{21}-2k_{11}k_{12}k_{21}k_{22}$,
$n_{2}=m_{22}k^{2}_{11}+m_{11}k^{2}_{21}-2m_{12}k_{11}k_{21}$,
$n_{3}=m_{22}k^{2}_{12}+m_{11}k^{2}_{22}-2m_{12}k_{12}k_{22}$ and
$n_{4}=m_{11}m_{22}-m_{12}^{2}$.
\end{theorem}

{\bf Proof.} Suppose that the $(1+1)$-mode Gaussian state $\rho_{AB}$ has CM $\Gamma_0=\left(\begin{array}{cccc}a & 0 & c & 0\\
0 & a & 0 & d\\
c & 0 & b & 0\\
0 & d & 0 & b\end{array}\right)$ and the mean $(\mathbf d_{A},
\mathbf d_{B})$. Then the CM $\Gamma^\prime$ and the mean $\mathbf
d^\prime$ of $\sigma_{AB}=(I\otimes \Phi)\rho_{AB}$ are respectively
\begin{align*}
&\Gamma^{\prime} = \left(\begin{array}{cc}
I & 0 \\
0 & K
\end{array}
\right)\left(\begin{array}{cc}
A_{0} & C_{0} \\
C^{T}_{0} & B_{0}
\end{array}
\right)\left(\begin{array}{cc}
I & 0 \\
0 & K^{T}
\end{array}
\right)+\left(\begin{array}{cc}
0 & 0 \\
0 & M
\end{array}
\right)=\left(\begin{array}{cc}
A_{0} & C_{0}K^{T} \\
KC_{0}^{T} & KB_{0}K^{T}+M
\end{array}
\right)
\end{align*}
and $$\mathbf d^{\prime}=(I\oplus K)(\mathbf d_{A}\oplus \mathbf
d_{B})+0\oplus \overline{\mathbf d}=\mathbf d_{A}\oplus (K\mathbf
d_{B}+\overline{\mathbf d}).$$

After a local invariant Gaussian unitary operation on the subsystem
A, one has $(U\otimes I)\sigma_{AB}(U^{\dag}\otimes
I)=\sigma_{AB}^{\prime}$. Remind that $U\rho_AU^\dag=\rho_A$, which
forces that, at the symplectic transformation level, $U=U_{\mathbf
S,\mathbf m}$ with $\mathbf m=0$ and ${\bf S} = {\bf S}_{\theta} =
\left(
\begin{array}{cc}
\cos\theta & \sin\theta \\
-\sin\theta & \cos\theta
\end{array}
\right)$ for some $\theta\in [0, \frac{\pi}{2}]$. Hence the CM and
the mean of $\sigma_{AB}^\prime$ are respectively
\begin{align*}
&\Gamma^{\prime} _{\bf S}= \left(\begin{array}{cc}
\bf S & 0 \\
0 & I
\end{array}
\right)\left(\begin{array}{cc}
A_{0} & C_{0}K^{T} \\
KC_{0}^{T} & KB_{0}K^{T}+M
\end{array}
\right)\left(\begin{array}{cc}
\bf S^{T} & 0 \\
0 & \textit{I}
\end{array}
\right)=\left(\begin{array}{cc}
 A_{0}  & {\bf S} C_{0}K^{T} \\
KC_{0}^{T} \bf S^{T} & KB_{0}K^{T}+M
\end{array}
\right)
\end{align*} and
$$\mathbf d^{\prime}_{\bf S}=({\bf S}\oplus I)(\mathbf d_{A}\oplus (K\mathbf
d_{B}+\overline{\mathbf d}))+\mathbf m \oplus 0=({\bf S} \mathbf
d_{A}+\mathbf m )\oplus (K\mathbf d_{B}+\overline{\mathbf
d})=\mathbf d_{A}\oplus (K\mathbf d_{B}+\overline{\mathbf d}).$$
  After some straight-forward calculations, one can
immediately get
\begin{align*}
&N_{\mathcal F}^{\mathcal G}((I\otimes \Phi)\rho_{AB}) =
N_{\mathcal F}^{\mathcal G}(\sigma_{AB}) \\
=& \sup_{U\in{\mathcal{U}^{\mathcal G}(H_A)},\
U\sigma_AU^\dag=\sigma_A} C^{2}(\sigma_{AB},
(U\otimes \textit{I})\sigma_{AB}(U^{\dag}\otimes \textit{I})) \\
 =& \sup_{\theta\in[0,\frac{\pi}{2}]}\{1-\frac{\sqrt{\det\Gamma^{\prime}\det\Gamma^{\prime}_{\bf S_\theta}}}{{\det((\Gamma^{\prime} +\Gamma^{\prime}_{\bf S_\theta})/2)}} \}.
\end{align*}
By the fact that $\det\Gamma^{\prime}=\det\Gamma^{\prime}_{\bf
S}=\det A_{0} \det(KB_{0}K^{T}+M-KC_{0}^{T}A_{0}^{-1}C_{0}K^{T})$,
the above formula can rewritten as the following
\begin{align*}
N_{\mathcal F}^{\mathcal G}((I\otimes \Phi)\rho_{AB})
 =& \sup_{\theta\in[0,\frac{\pi}{2}]}\{1-\frac{\det \left(\begin{array}{cc}A_{0} & C_{0}K^{T} \\
KC_{0}^{T} & KB_{0}K^{T}+M\end{array}\right)}{\det \left(\begin{array}{cc}A_{0}  & \frac{(I+{\bf S_\theta}) C_{0}K^{T}}{2} \\
\frac{KC_{0}^{T}(I+{\bf S}_\theta^{T})}{2} &
KB_{0}K^{T}+M\end{array}\right)}
\} \\
=&\sup_{\theta\in[0,\frac{\pi}{2}]}\{1-\frac{\det A_{0}
\det(KB_{0}K^{T}+M-KC_{0}^{T}A_{0}^{-1}C_{0}K^{T})}{\det A_{0}
\det(KB_{0}K^{T}+M-\frac{KC_{0}^{T}(I+{\bf
S}_\theta^{T})}{2}A_{0}^{-1}\frac{(I+{\bf S}_\theta) C_{0}K^{T}}{2})}\}\\
=& \sup_{\theta\in[0,\frac{\pi}{2}]}\{1-\frac{
\det(K(B_{0}-C_{0}^{T}A_{0}^{-1}C_{0})K^{T}+M)}{
\det(K(B_{0}-\frac{C_{0}^{T}(I+{\bf
S}_\theta^{T})}{2}A_{0}^{-1}\frac{(I+{\bf S}_\theta)
C_{0}}{2})K^{T}+M)}\}.
\end{align*}
Clearly,  the quantity $N_{\mathcal F}^{\mathcal G}((I\otimes
\Phi)\rho_{AB})$ is independent of the parameter $\overline{\mathbf
d}$. Notice that $K$, $M$ can not be zero simultaneously,
substituting ${\bf S}_{\theta} = \left(
\begin{array}{cc}
\cos\theta & \sin\theta \\
-\sin\theta & \cos\theta
\end{array}
\right)$   into the above equation, after tedious calculations, one
has
\begin{eqnarray*}
&&N_{\mathcal F}^{\mathcal G}((I\otimes \Phi)\rho_{AB}) \\
&=&\sup_{\theta\in[0,\frac{\pi}{2}]}\{1-\frac{
\det(K(B_{0}-C_{0}^{T}A_{0}^{-1}C_{0})K^{T}+M)}{
\det(K(B_{0}-\frac{C_{0}^{T}(I+{\bf
S}_\theta^{T})}{2}A_{0}^{-1}\frac{(I+{\bf
S}_\theta) C_{0}}{2})K^{T}+M)}\}\\
&=&
\sup_{\theta\in[0,\frac{\pi}{2}]}\{1-\frac{(ab-c^{2})(ab-d^{2})n_{1}+a(ab-c^{2})n_{2}+a(ab-d^{2})n_{3}+a^{2}n_{4}}
{(ab-c^{2}n_{0})(ab-d^{2}n_{0})n_{1}+a(ab-c^{2}n_{0})n_{2}+a(ab-d^{2}n_{0})n_{3}+a^{2}n_{4}}\}\\
&=&1-\frac{(ab-c^{2})(ab-d^{2})n_{1}+a(ab-c^{2})n_{2}+a(ab-d^{2})n_{3}+a^{2}n_{4}}
{(ab-c^{2}/2)(ab-d^{2}/2)n_{1}+a(ab-c^{2}/2)n_{2}+a(ab-d^{2}/2)n_{3}+a^{2}n_{4}},
\end{eqnarray*}
where \begin{align*}  n_{0}=& (1+\cos\theta)/2,\\
 n_{1}=&
 k^{2}_{11}k^{2}_{22}+k^{2}_{12}k^{2}_{21}-2k_{11}k_{12}k_{21}k_{22},
 &
 n_{2}=& m_{22}k^{2}_{11}+m_{11}k^{2}_{21}-2m_{12}k_{11}k_{21},\\
 n_{3}=& m_{22}k^{2}_{12}+m_{11}k^{2}_{22}-2m_{12}k_{12}k_{22}, &
 n_{4}=& m_{11}m_{22}-m_{12}^{2}. \end{align*}
The proof is completed. \hfill$\Box$

{\bf Remark 2.} If $K=0$, then $\det M\geq 1$, and we have
\begin{align*}
N_{\mathcal F}^{\mathcal G}((I\otimes \Phi(0, M, \overline{\mathbf
d}))\rho_{AB}) =&\{1-\frac{ \det M }{ \det M}\} =0.
\end{align*}
In fact, in this case, the Gaussian channel $I\otimes \Phi(0, M,
\overline{\mathbf d})$ maps any Gaussian state $\rho_{AB}$ to a
product state. Thus, by Theorem 3, we always have $N_{\mathcal
F}^{\mathcal G}((I\otimes \Phi(0, M, \overline{\mathbf
d}))\rho_{AB})=0$.

{\bf Remark 3.} If $M=0$, then $\det K=1=\det K^{T}$, and
\begin{align*}
N_{\mathcal F}^{\mathcal G}((I\otimes \Phi(K, 0, \overline{\mathbf
d}))\rho_{AB})=& \sup_{\theta\in[0,\frac{\pi}{2}]}\{1-\frac{
\det(K(B_{0}-C_{0}^{T}A_{0}^{-1}C_{0})K^{T})}{
\det(K(B_{0}-\frac{C_{0}^{T}(I+{\bf
S}_\theta^{T})}{2}A_{0}^{-1}\frac{(I+{\bf S}_\theta)
C_{0}}{2})K^{T})}\}
\\=& \sup_{\theta\in[0,\frac{\pi}{2}]}\{1-\frac{ \det(B_{0}-C_{0}^{T}A_{0}^{-1}C_{0})}{
\det(B_{0}-\frac{C_{0}^{T}(I+{\bf
S}_\theta^{T})}{2}A_{0}^{-1}\frac{(I+{\bf
S}_\theta) C_{0}}{2})}\}\\
=& \sup_{\theta\in[0,\frac{\pi}{2}]}\{1-\frac{\det \left(\begin{array}{cc}A_{0} & C_{0} \\
C_{0}^{T} & B_{0}\end{array}\right)}{\det \left(\begin{array}{cc}A_{0}  & \frac{(I+{\bf S}_\theta) C_{0}}{2} \\
\frac{C_{0}^{T}(I+{\bf S}_\theta^{T})}{2} & B_{0}\end{array}\right)}
\} = N_{\mathcal F}^{\mathcal G}(\rho_{AB}).
\end{align*}
In this case, one can conclude that, after performing the Gaussian
operation $I\otimes \Phi(K, 0, \overline{\mathbf d})$, the quantity
 $N_{\mathcal F}^{\mathcal G}$ remains the same for those
$(1+1)$-mode Gaussian states whose CM are of the standard form.

The following result gives a kind of {\it local Gaussian operation
non-increasing property} of $N_{\mathcal F}^{\mathcal G}$, which was
not discussed for  other known similar correlations such as the
Gaussian QD (two-mode) \cite{Giorda, Adesso}, Gaussian geometric
discord \cite{Adesso-Girolami}, the correlations $Q$, $Q_{\mathcal
P}$ discussed in \cite{MHQ} and the correlation ${\mathcal N}$
discussed in \cite{WHQ}.

\begin{theorem}
Let $\rho_{AB}$ be a $(1+1)$-mode Gaussian state. Then, for any
Gaussian channel $\Phi$   performed on the subsystem $B$, we have
\begin{align*}
0\leq N_{\mathcal F}^{\mathcal G}((I\otimes \Phi)\rho_{AB})\leq
N_{\mathcal F}^{\mathcal G}(\rho_{AB}).
\end{align*}
\end{theorem}

{\bf Proof.}  We first consider the case that  the $(1+1)$-mode
Gaussian states $\rho_{AB}$ whose CM $\Gamma_0$ are of the standard
form,
that is, $\Gamma_0=\left(\begin{array}{cccc}a & 0 & c & 0\\
0 & a & 0 & d\\
c & 0 & b & 0\\
0 & d & 0 & b\end{array}\right).$ Let $\Phi=\Phi(K, M,
\overline{\mathbf d})$ be any  Gaussian channel performed on subsystem B with $K=\left(\begin{array}{cc}k_{11}&k_{12}\\
k_{21} & k_{22}\end{array}\right)$ and $M=\left(\begin{array}{cc}m_{11} &m_{12}\\
m_{12} & m_{22}\end{array}\right).$ We  have to show that
$N_{\mathcal F}^{\mathcal G}((I\otimes \Phi)\rho_{AB}) \leq
N_{\mathcal F}^{\mathcal G}(\rho_{AB})$.

If $N_{\mathcal F}^{\mathcal G}(\rho_{AB})=0$, then, by Theorem 3,
$\rho_{AB}$ is a product state. So $(I\otimes \Phi)\rho_{AB}$ is a
product state, and hence $N_{\mathcal F}^{\mathcal G}((I\otimes
\Phi)\rho_{AB})=0=N_{\mathcal F}^{\mathcal G}(\rho_{AB})$.

Assume that $N_{\mathcal F}^{\mathcal G}(\rho_{AB})\neq 0$. Then
$N_{\mathcal F}^{\mathcal G}((I\otimes \Phi)\rho_{AB}) \leq
N_{\mathcal F}^{\mathcal G}(\rho_{AB})$ holds if and only if
$\frac{N_{\mathcal F}^{\mathcal G}((I\otimes
\Phi)\rho_{AB})}{N_{\mathcal F}^{\mathcal G}(\rho_{AB})}\leq 1$. Let
$\alpha=(ab-c^{2})(ab-d^{2})$, $\beta=(ab-c^{2}/2)(ab-d^{2}/2)$,
$\gamma=a(ab-c^{2})n_{2}+a(ab-d^{2})n_{3}+a^{2}n_{4}$ and
$\delta=a(ab-c^{2}/2)n_{2}+a(ab-d^{2}/2)n_{3}+a^{2}n_{4}$ with
$n_2,n_3, n_4$ as in Theorem 6.  Then, according to Theorem 6, we
have
$$\frac{N_{\mathcal F}^{\mathcal G}((I\otimes \Phi)\rho_{AB})}{N_{\mathcal F}^{\mathcal G}(\rho_{AB})}\leq1 \Leftrightarrow \frac{1-\frac{\alpha n_{1}+\gamma}{\beta
n_{1}+\delta}}{1-\frac{\alpha}{\beta}}\leq 1\Leftrightarrow
\frac{\alpha n_{1}+\gamma}{\beta n_{1}+\delta}\geq
\frac{\alpha}{\beta}\Leftrightarrow \gamma\beta\geq \alpha\delta. $$
Therefore, it suffices to prove that $\gamma\beta-\alpha\delta
\geq0$. By some computations, one sees that
\begin{eqnarray*}
&&\gamma\beta =[a(ab-c^{2})n_{2}+a(ab-d^{2})n_{3}+a^{2}n_{4}](ab-\frac{c^{2}}{2})(ab-\frac{d^{2}}{2})\\
&=&
a(ab-c^{2})(ab-\frac{c^{2}}{2})(ab-\frac{d^{2}}{2})n_{2}+a(ab-d^{2})(ab-\frac{c^{2}}{2})(ab-\frac{d^{2}}{2})n_{3}+a^{2}(ab-\frac{c^{2}}{2})(ab-\frac{d^{2}}{2})n_{4}
\end{eqnarray*}
and
\begin{eqnarray*}
&&\alpha\delta =
a(ab-c^{2})(ab-\frac{c^{2}}{2})(ab-d^{2})n_{2}+a(ab-d^{2})(ab-c^{2})(ab-\frac{d^{2}}{2})n_{3}+a^{2}(ab-c^{2})(ab-d^{2})n_{4}.
\end{eqnarray*}
Note that
$n_{1}=k^{2}_{11}k^{2}_{22}+k^{2}_{12}k^{2}_{21}-2k_{11}k_{12}k_{21}k_{22}=(k_{11}k_{22}-k_{12}k_{21})^{2}\geq
0$ and $n_{4}=m_{11}m_{22}-m_{12}^{2}=\det M\geq 0$. Since
$m_{22}k^{2}_{11}+m_{11}k^{2}_{21}\geq 2 \sqrt m_{22} \sqrt
m_{11}k_{11}k_{21}\geq 2 m_{12}k_{11}k_{21}$, we have $n_{2}\geq 0$.
One can verify $n_{3}\geq 0$ by the same way.
 Also note that $a,b\geq 1$
and $ab\geq c ^2(d^2)$ by the constraint condition of the parameters
in the definition of the Gaussian state. Now it is clear that
\begin{eqnarray*}
&&\gamma\beta-\alpha\delta =
a(ab-c^{2})(ab-\frac{c^{2}}{2})\frac{d^{2}}{2}n_{2}+a(ab-d^{2})(ab-\frac{d^{2}}{2})\frac{c^{2}}{2}n_{3}+a^{2}\frac{c^{2}}{2}\frac{d^{2}}{2}n_{4}\geq
0,
\end{eqnarray*}
as desired. To this end, we come to the conclusion that $N_{\mathcal
F}^{\mathcal G}((I\otimes \Phi)\rho_{AB}) \leq N_{\mathcal
F}^{\mathcal G}(\rho_{AB})$, and the equality holds if $M=0$ (See
Remark 3 after the proof of Theorem 6).

Now let us consider the general case. Let ${\mathcal U}\otimes
{\mathcal V}$ be a local Gaussian unitary operation, that is, for
some Gaussian unitary operators $U$ and $V$ on the subsystem A and B
respectively, so that $({\mathcal U}\otimes {\mathcal
V})(\rho_{AB})=(U\otimes V)\rho_{AB}(U^\dag\otimes V^\dag)$ for each
state $\rho_{AB}$. Then, $$(I\otimes \Phi)\circ({\mathcal U}\otimes
{\mathcal V})={\mathcal U}\otimes (\Phi\circ{\mathcal V})=({\mathcal
U}\otimes I)\circ(I\otimes (\Phi\circ{\mathcal V})).$$ Note that,
$\Phi\circ{\mathcal V}$ is still a Gaussian channel which sends
$\rho_B$ to $\Phi(V\rho_BV^\dag)$. Keep this in mind and let
$\rho_{AB}$ be any $(1+1)$-mode Gaussian state. Then there exists a
local Gaussian unitary operation $U\otimes V$ such that
$\sigma_{AB}=(U^\dag\otimes V^\dag)\rho_{AB}(U\otimes V)$ has CM of
the standard form. By what we have proved above and Theorem 2, we
see that
\begin{align*}
 N_{\mathcal F}^{\mathcal G}((I\otimes \Phi)\rho_{AB})=& N_{\mathcal
F}^{\mathcal G}((I\otimes \Phi)((U\otimes
V)\sigma_{AB}(U^\dag\otimes
V^\dag)))\\
=& N_{\mathcal F}^{\mathcal G}((I\otimes \Phi)\circ({\mathcal
U}\otimes {\mathcal V})\sigma_{AB})=N_{\mathcal F}^{\mathcal
G}(({\mathcal U}\otimes I)\circ(I\otimes (\Phi \circ  {\mathcal
V}))\sigma_{AB})\\
=&N_{\mathcal F}^{\mathcal G}((I\otimes (\Phi \circ  {\mathcal
V}))\sigma_{AB})\leq N_{\mathcal F}^{\mathcal G}(\sigma_{AB})=
N_{\mathcal F}^{\mathcal G}(\rho_{AB}),
\end{align*}
as desired, which completes the proof. \hfill$\Box$

\section{Comparison between $N_{\mathcal F}^{\mathcal
G}$ and other quantifications of the Gaussian quantum correlations}

$N_{\mathcal F}^{\mathcal G}$, $D_{G}$ and $Q$ describe the same
quantum nonclassicality  when they are restricted to Gaussian states
because they take value 0 at a Gaussian state $\rho_{AB}$  if and
only if $\rho_{AB}$ is a product state.  In this section, we
calculate $N_{\mathcal F}^{\mathcal G}(\rho_{AB})$ for all two-mode
symmetric squeezed thermal states $\rho_{AB}$ and compare it with
Gaussian geometric discord $D_{G}(\rho_{AB})$ and $Q(\rho_{AB})$ in
scale. Our result reveals that $N_{\mathcal F}^{\mathcal G}$ is
bigger and thus is easier to detect the correlation in states. Since
the known computation formula of $D_{G}(\rho_{AB})$ is only for
symmetric squeezed thermal states $\rho_{AB}$, we compare them on
such states.

{\it Symmetric squeezed thermal states: } Assume that $\rho_{AB}$ is
any two-mode Gaussian state; then its standard CM has the form as in
Eq.(3). Recall that the symmetric squeezed thermal states (SSTSs)
are Gaussian states whose CMs are parameterized by $\bar{n}$ and
$\mu$ such that $a=b=1+2\bar{n}$ and
$c=-d=2\mu\sqrt{\bar{n}(1+\bar{n})}$, where $\bar{n}$ is the mean
photon number for each part and $\mu$ is the mixing parameter with
$0\leq \mu\leq 1$ (ref. \cite{WRPT}). Thus every SSTS may be denoted
by $\rho_{AB}(\bar{n},\mu)$.

Thus by Theorem 4, for any SSTS $\rho_{AB}(\bar{n},\mu)$, we have
\begin{align} N_{\mathcal F}^{\mathcal G}(\rho_{AB}(\bar{n},\mu))=1-\frac{((1+2\bar
n)^{2}-4  \mu^{2} \bar n(1+\bar n))^{2}}{((1+2\bar n)^{2}-2  \mu^{2}
\bar n(1+\bar n))^{2}}.\end{align}

For any two-mode Gaussian state $\rho_{AB}$, recall that the
Gaussian geometric discord of $\rho_{AB}$ (\cite{Adesso-Girolami})
is defined as
$$D_{G}(\rho_{AB})=\inf_{\Pi^{A}}\|\rho_{AB}-\Pi^{A}(\rho_{AB})\|^{2}_{2},$$
where $\Pi^{A}={\Pi^{A}(\alpha)}$ runs over all Gaussian positive
operator valued measurements of subsystem A,
$\Pi^{A}(\rho_{AB})=\int(\Pi^{A}(\alpha)\otimes I)^{\frac{1}{2}}
\rho_{AB} (\Pi^{A}(\alpha)\otimes I)^{\frac{1}{2}} {\rm d}^{2}
\alpha $. According to the analytical formula of $D_{G}(\rho_{AB})$
provided in \cite{Adesso-Girolami}, for any SSTS $\rho_{AB}$ with
parameters $\bar{n}$ and $\mu$, one has
\begin{align}D_{G}(\rho_{AB}(\bar{n},\mu))=\frac{1}{(1+2\bar{n})^{2}-4\mu^{2}\bar{n}(1+\bar{n})}
-\frac{9}{[\sqrt{4(1+2\bar{n})^{2}-12\mu^{2}\bar{n}(1+\bar{n})}+(1+2\bar{n})]^{2}}.\end{align}

By Eqs.(9)-(10), it is clear that $$\lim _{\bar{n}\to
\infty}N_{\mathcal F}^{\mathcal
G}(\rho_{AB}(\bar{n},\mu))=1-\frac{(1-\mu^2)^2}{(1-\frac{1}{2}\mu^2)^2}>0
\quad {\rm for}\ \mu\in(0,1),$$ while $$ \lim _{\bar{n}\to
\infty}D_{G}(\rho_{AB}(\bar{n},\mu))=0 \quad {\rm for}\
\mu\in(0,1).$$ This shows that, for the case $\mu\not=0, 1$,
$N_{\mathcal F}^{\mathcal G}$ is able to recognize well the quantum
correlation in the states with large mean photon number but $D_{G}$
is not. It is clear that $\mu=0$ if and only if $\rho_{AB} $ is a
product SSTS, and in this case, $N_{\mathcal F}^{\mathcal
G}(\rho_{AB}(\bar{n},0))=D_{G}(\rho_{AB}(\bar{n},0))=0$. When
$\mu=1$, we have
$$ N_{\mathcal F}^{\mathcal G}(\rho_{AB}(\bar{n},1))=1-\frac{1}{(1+2\bar n +2 \bar{n}^{2})^{2}}$$
and
$$D_{G}(\rho_{AB}(\bar{n},1))=1
-\frac{9}{[1+2\bar{n}+2\sqrt{1+\bar{n}+ \bar{n}^2}]^{2}},$$ which
reveals that we always have
$$N_{\mathcal F}^{\mathcal
G}(\rho_{AB}(\bar{n},1))>D_{G}(\rho_{AB}(\bar{n},1)).$$ Moreover, we
randomly chose 100000 pairs of $(\bar n,\mu)$ with $\bar
n\in(0,10000000000000)$ and $\mu\in(0,1)$,
 numerical results show that $N_{\mathcal F}^{\mathcal G}(\rho_{AB}(\bar{n},\mu) )> D_{G}(\rho_{AB}(\bar{n},\mu))$.
 On the other hand, the numerical method suggests that $N_{\mathcal
F}^{\mathcal G}$ is better than $D_G$ in detecting the quantum
correlation contained in any SSTS because we always have
$$N_{\mathcal F}^{\mathcal G}(\rho_{AB}(\bar{n},\mu) )> D_{G}(\rho_{AB}(\bar{n},\mu))$$ for all
SSTSs $\rho_{AB}(\bar{n},\mu)$ with $\mu\not=0$.

In Fig.1, we compare $N_{\mathcal F}^{\mathcal G}(\rho_{AB})$ with
$D_{G}(\rho_{AB})$ for SSTSs $\rho_{AB}$ by considering $N_{\mathcal
F}^{\mathcal G}(\rho_{AB})-D_{G}(\rho_{AB})$ for $\bar{n}\leq 50$.
Fig.1 shows that $N_{\mathcal F}^{\mathcal
G}(\rho_{AB})-D_{G}(\rho_{AB})\geq 0$ and
$$N_{\mathcal F}^{\mathcal G}(\rho_{AB})\gg D_{G}(\rho_{AB})$$ for
SSTSs $\rho_{AB}$ with $\mu$ near 1. For example,
 considering the state $\rho_{AB}$ with
$\bar{n}=49$ and $\mu=0.9$, we have  $D_{G}(\rho_{AB})\approx
0.000356$, which is very close to 0 and difficult to judge weather
or not $\rho_{AB}$ contains the correlation. However, $N_{\mathcal
F}^{\mathcal G}(\rho_{AB})\approx 0.897995\gg 0$, which guarantees
that $\rho_{AB}$ does contain the quantum correlation. For large
mean photon number, for example, $\bar{n}=10000$, taking $\mu=0.9$,
we have $N_{\mathcal F}^{\mathcal G}(\rho_{AB})\approx 0.89803\gg
0$, but $D_{G}(\rho_{AB})\approx 8.72518 \times 10^{-11}$.
Furthermore,
 Fig.2 shows that
$N_{\mathcal F}^{\mathcal G}(\rho_{AB})-D_{G}(\rho_{AB})\geq 0$
holds as well for $\bar n\in(100000,100500)$ and $\mu\in(0,1)$.

$Q$ is a quantum correlation for $(m+n)$-mode continuous-variable
systems defined in terms of average distance between the reduced
states under the local Gaussian positive operator valued
measurements \cite{MHQ}:
\begin{eqnarray*}
Q(\rho_{AB}):=\sup\limits_{\Pi^{A}}\int
p(\alpha){\|\rho_{B}-\rho^{(\alpha)}_{B}\|^{2}_{2}}{\rm
d}^{2m}\alpha,
\end{eqnarray*}
where $\Pi^{A}={\Pi^{A}(\alpha)}$ runs over all Gaussian positive
operator valued measurements of subsystem A,
$\Pi^{A}=\{\Pi^{A}(\alpha)\}$ on the subsystem $H_A$, $\rho_{B}={\rm
Tr}_{A}(\rho_{AB})$, $p(\alpha)={\rm Tr}[(\Pi^{A}(\alpha)\otimes
I_{B})\rho_{AB}]$ and $\rho^{(\alpha)}_{B}=\frac{1}{p(\alpha)}{\rm
Tr}_{A}[(\Pi^{A}(\alpha)\otimes
I_{B})^{\frac{1}{2}}\rho_{AB}(\Pi^{A}(\alpha)\otimes
I_{B})^{\frac{1}{2}}]$.

For any SSTS $\rho_{AB}$ with parameters $\bar{n}$ and $\mu$, by
\cite{MHQ}, \begin{align} Q(\rho_{AB}(\bar{n},\mu)
)=\frac{1}{1+2\bar {n}(1-\mu^{2})}-\frac{1}{1+2\bar {n}}.\end{align}
Obviously,
$$\lim_{\bar{n}\to\infty} Q(\rho_{AB}(\bar{n},\mu) )=0,\quad{\rm for}\ \mu\in(0,1).$$
which reveals that $Q$ is not valid for those states with
$\mu\in(0,1)$ and large mean photon number. For the case $\mu=1$, we
have
$$Q(\rho_{AB}(\bar{n},1)
)=1-\frac{1}{1+2\bar {n}}<N_{\mathcal F}^{\mathcal
G}(\rho_{AB}(\bar{n},1))
$$
for any $\bar{n}$. Also, we always have
$$N_{\mathcal F}^{\mathcal G}(\rho_{AB})>Q(\rho_{AB})$$ for all
SSTSs with $\mu\not=0$. For random pairs $(\bar n,\mu)$ with $\bar
n\in(0,10000000000000)$ and $\mu\in(0,1)$, 100000 numerical results
illustrate that $N_{\mathcal F}^{\mathcal G}(\rho_{AB}(\bar{n},\mu)
)> Q(\rho_{AB}(\bar{n},\mu))$.

\begin{figure}
\includegraphics[width=0.6\linewidth]{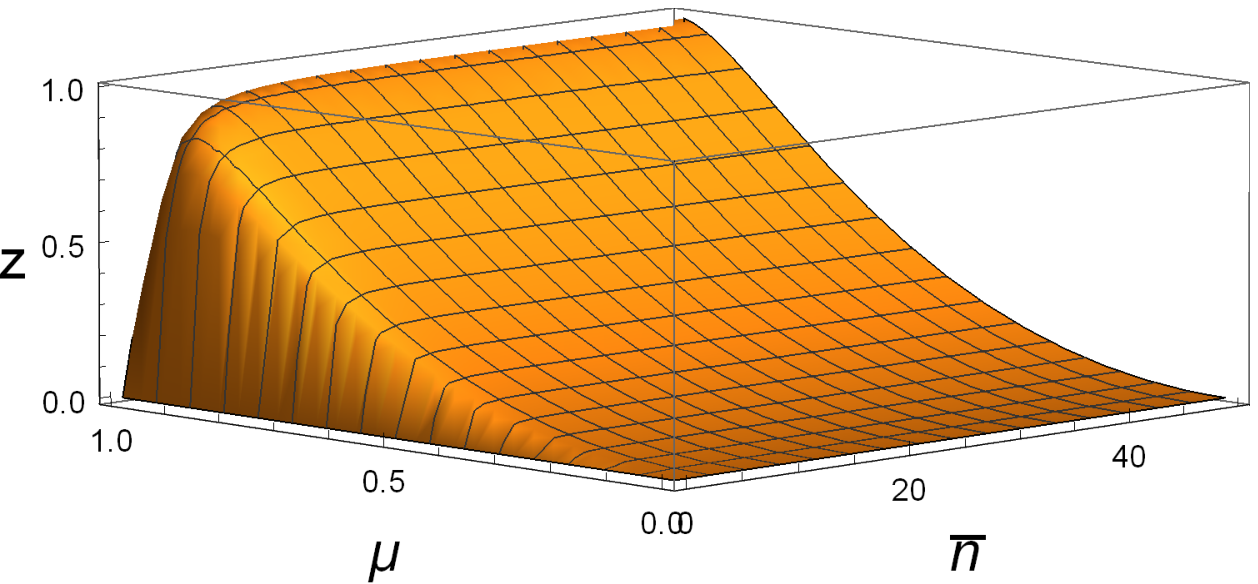}
\caption{z=$N_{\mathcal F}^{\mathcal G}(\rho_{AB})-D_{G}(\rho_{AB})$
with SSTSs, and $0\leq\mu\leq 1$, $0\leq \bar{n} \leq 50$.}
\end{figure}
\begin{figure}
\includegraphics[width=0.6\linewidth]{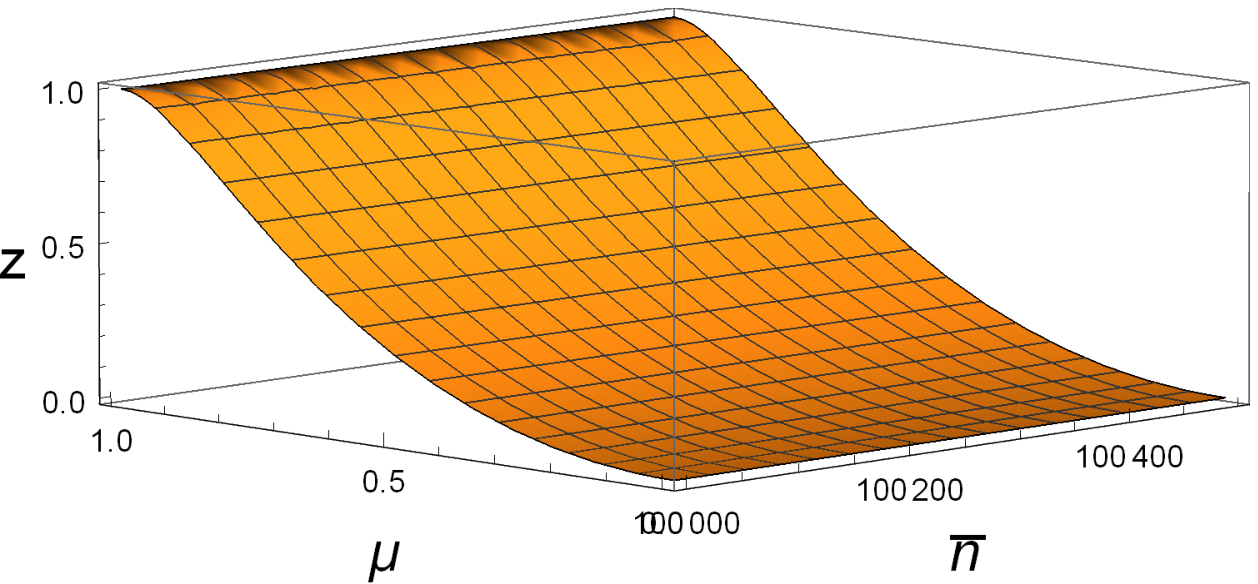}
\caption{z=$N_{\mathcal F}^{\mathcal G}(\rho_{AB})-D_{G}(\rho_{AB})$
with SSTSs, and $0\leq\mu\leq 1$, $100000\leq \bar{n} \leq100500$.}
\end{figure}
\begin{figure}
\includegraphics[width=0.6\linewidth]{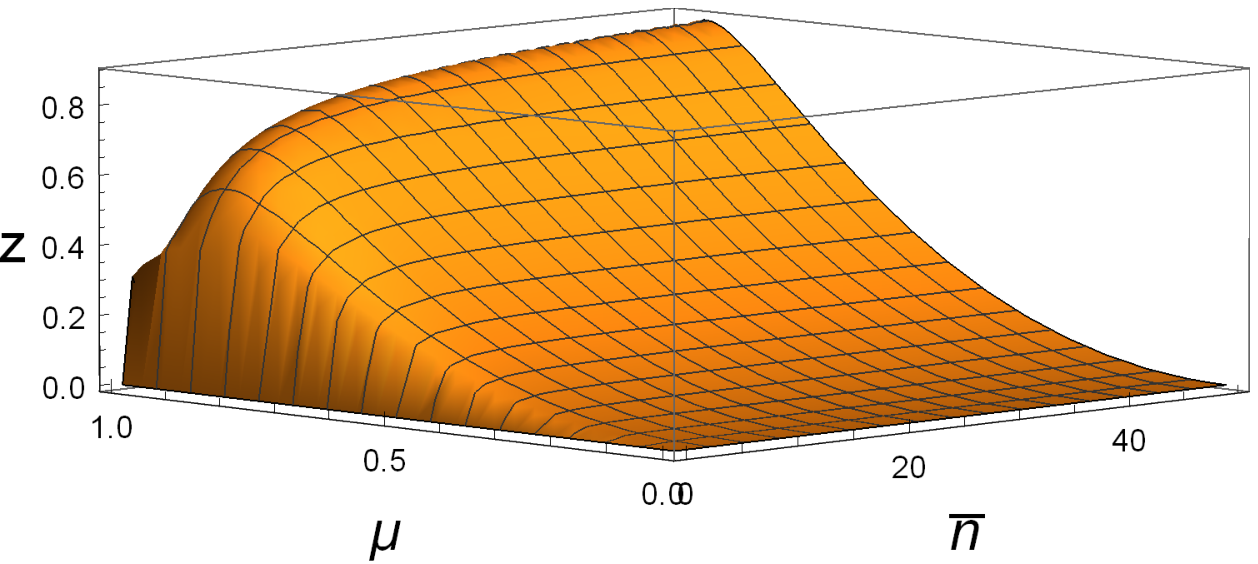}
\caption{z=$N_{\mathcal F}^{\mathcal G}(\rho_{AB})-Q(\rho_{AB})$
with SSTSs, and $0\leq\mu\leq 1$, $0\leq \bar{n} \leq 50$.}
\end{figure}
\begin{figure}
\includegraphics[width=0.6\linewidth]{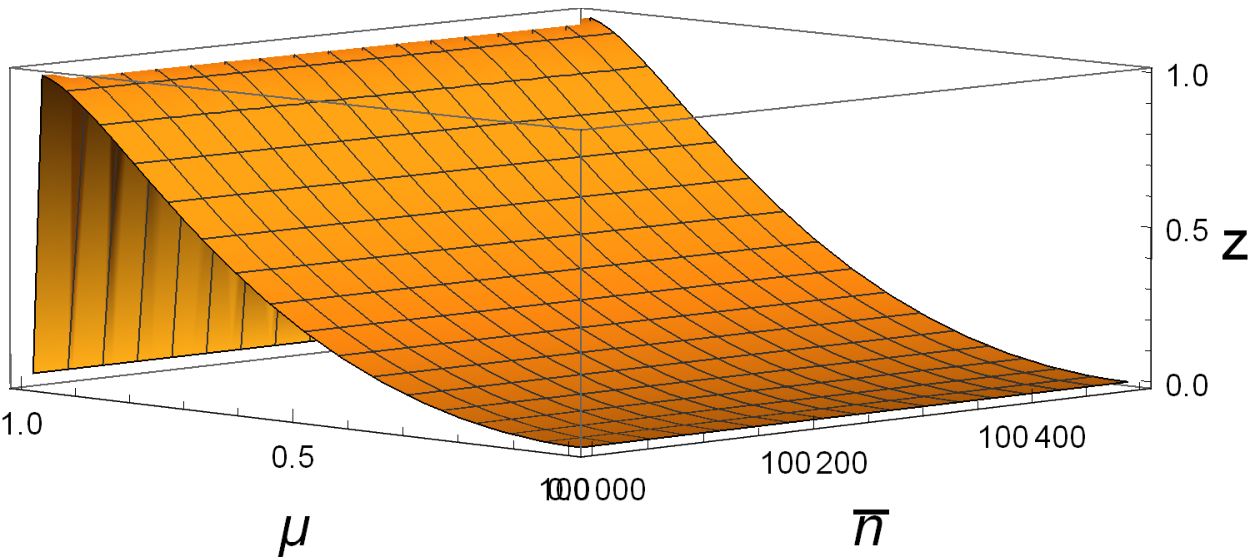}
\caption{z=$N_{\mathcal F}^{\mathcal G}(\rho_{AB})-D_{G}(\rho_{AB})$
with SSTSs, and $0\leq\mu\leq 1$, $100000\leq \bar{n} \leq100500$.}
\end{figure}

The deference of $N_{\mathcal F}^{\mathcal G}(\rho_{AB})$ and
$Q(\rho_{AB})$ for SSTSs is showed in Fig.3 for $\bar{n}\leq 50$. It
reveals that $N_{\mathcal F}^{\mathcal G}(\rho_{AB})\gg
Q(\rho_{AB})$ for those SSTSs $\rho_{AB}$ with large mean photon
number $\bar{n}$ and larger mixing parameter $\mu$. Consider the
states $\rho_{AB}$ with respectively $(\bar{n}, \mu)=(49, 0.9)$ and
$(\bar{n}, \mu)=(10000, 0.9)$, the same examples as above. We have
respectively $Q(\rho_{AB})\approx 0.040867<N_{\mathcal F}^{\mathcal
G}(\rho_{AB})\approx 0.897955$ and $Q(\rho_{AB})\approx 0.000021\ll
N_{\mathcal F}^{\mathcal G}(\rho_{AB})\approx 0.89803$, which means
that applying $N_{\mathcal F}^{\mathcal G}$ is much more easier than
$Q$ to guarantee that $\rho_{AB}$ contains the quantum correlation.
Fig.4 demonstrates that $N_{\mathcal F}^{\mathcal
G}(\rho_{AB})-Q(\rho_{AB})\geq 0$ also holds for these $\bar
n\in(100000,100500)$ and $\mu\in(0,1)$.

\section{Conclusion}

In this paper, based on fidelity ${\mathcal
F}(\rho,\sigma)=\frac{|{\rm tr}\rho\sigma|}{\sqrt{{\rm tr}\rho^2{\rm
tr}\sigma^2}}$ and the distance $C^2(\rho,\sigma)=1-{\mathcal
F}^2(\rho,\sigma)$, we have proposed a new kind of quantum
nonclassicality $N_{\mathcal F}^{\mathcal G}$ by local Guassian
unitary operations for any states in $(n+m)$-mode
continuous-variable systems. Though, when restricted to the Gaussian
states, $N_{\mathcal F}^{\mathcal G}$ describes the same
nonclassical correlation as several known correlations such as
Gaussian QD, Gaussian geometric discord $D_G$ and the nonlocality
$Q$,  it is comparatively much easier to be computed and estimated.
Furthermore, $N_{\mathcal F}^{\mathcal G}$ has several nice
properties that other known quantifications of such correlation do
not possess: $N_{\mathcal F}^{\mathcal G}$ is a quantum correlation
without ancilla problem; $N_{\mathcal F}^{\mathcal G}((I\otimes
\Phi)\rho_{AB}) \leq N_{\mathcal F}^{\mathcal G}(\rho_{AB})$ holds
for any $(1+1)$-mode Gaussian state $\rho_{AB}$ and any Gaussian
channel $\Phi$, that is, undergoing a local Gaussian channel
performed on the unmeasured part, the quantity we proposed will not
increase. We guess that this nice property is still valid for
$(n+m)$-mode systems.  We give a computation formula of $N_{\mathcal
F}^{\mathcal G}$ for any $(1+1)$-mode Gaussian states and an upper
bound for any $(n+m)$-mode Gaussian states, which are simple and
easily calculated. Furthermore, by comparing $N_{\mathcal
F}^{\mathcal G}(\rho_{AB})$ with  $D_G(\rho_{AB})$ and
$Q(\rho_{AB})$ for two-mode symmetric squeezed thermal states, we
find  that $N_{\mathcal F}^{\mathcal G}$ is greater than $D_G$ and
$Q$, and so is better in detecting quantum correlation in Gaussian
states.

{\bf Acknowledgement.}  Authors thank Professor Kan He for
discussion. This work is supported by the National Natural Science
Foundation of China (11671294,11671006) and Outstanding Youth
Foundation of Shanxi Province (201701D211001).


\begin{thebibliography}{99}

\bibitem{Horodecki} R. Horodecki, P. Horodecki, M. Horodecki,  K. Horodecki, Rev. Mod. Phys. 81, 865 (2009).

\bibitem{Ollivier} H. Ollivier, W. H. Zurek, Phys. Rev. Lett. 88, 017901 (2001).

\bibitem{Borivoje} B. Daki\'{c}, V. Vedral,  \v{C}. Brukner, Phys. Rev. Lett. 105, 190502 (2010).

\bibitem{Luo} S. Luo, S. Fu, Phys. Rev. A 82, 034302 (2010).

\bibitem{Miranowicz} A. Miranowicz, P. Horodecki,  R. W. Chhajlany, et.al., Phys. Rev. A 86, 042123 (2012).

\bibitem{Luo-Fu} S. Luo, S. Fu, Phys. Rev. Lett. 106, 120401 (2011).

\bibitem{Luo-S} S. Luo, Phys. Rev. A 77, 022301 (2008).

\bibitem{Giorda} P. Giorda, M. G. A. Paris, Phys. Rev. Lett. 105, 020503
(2010).

\bibitem{Adesso} G. Adesso, A. Datta, Phys. Rev. Lett. 105, 030501 (2010).

\bibitem{Adesso-Girolami} G. Adesso, D. Girolami, Int. J. Quantum Inf. 09, 1773-1786 (2011).

\bibitem{Mista} L. Mi\v{s}ta, Jr., R. Tatham, D. Girolami, N. Korolkova, G. Adesso, Phys. Rev. A 83, 042325 (2011).

\bibitem{Ma} R.-F. Ma, J.-C. Hou, X.-F. Qi, Int. J. Theor. Phys. 56, 1132-1140 (2017).

\bibitem{MHQ} R.-F. Ma, J.-C. Hou, X.-F. Qi,  Y.-Y. Wang, Quantum Inf. Process.  17:98
(2018).

\bibitem{Farace} A. Farace, A. De. Pasquale1, L. Rigovacca, V.
Giovannetti, New J. Phys. 16, 073010 (2014).

\bibitem{Rigovacca} L. Rigovacca, A. Farace, A. D. Pasquale,  V. Giovannetti, Phys. Rev. A 92, 042331 (2015).

\bibitem{WHQ} Y.-Y. Wang, J.-C. Hou, X.-F. Qi, Entropy 20, 266
(2018).

\bibitem{Fu} L. Fu, Europhys. Lett. 75, 1 (2006).

\bibitem{Datta} A. Datta, S. Gharibian, Phys. Rev. A 79, 042325 (2009).




\bibitem{Gharibian} S. Gharibian, Phys. Rev. A 86, 042106 (2012).

\bibitem{Huang} Y. Huang. New J. Phys. 16, 033027 (2014).

\bibitem{CSYu} X. Wang, C.-S. Yu, X.-X. Yi, Phys. Lett. A 373, 58-60 (2008).

\bibitem{Braunstein} S. L. Braunstein, P. van Loock, Rev. Mod. Phys. 77, 513 (2005).


\bibitem{Wang} X.-B. Wang, T. Hiroshimab, A. Tomitab,  M. Hayashi, Phys. Rep. 448, 1-111 (2007).

\bibitem{Christian} C. Weedbrook, S. Pirandola, et al., Rev. Mod. Phys. 84, 621 (2012).

\bibitem{JA} J. Anders, arXiv:quant-ph/0610263(2006).



\bibitem{Holder} H. Lutkenpohl, Handbook of Matrices. John Wiley and son's Ltd, Chichester.
(1996).

\bibitem{Horn} R. A. Horn, C. R. Johnson, Matrix Analysis, Cambridge university Press, Cambridge, United Kingdom
(2012).


\bibitem{Jozsa} R. Jozsa. Journal of Mordern Optics, 41, 2315 (1994).

\bibitem{Massar} N. Gisin, S. Massar, Phys. Rev. Lett. 79, 2153-2156 (1997).

\bibitem{GFZhang} G.-F. Zhang, Phys. Rev. A 75, 034304 (2007).

\bibitem{Brida} Yu. I. Bogdanov, G. Brida, M. Genovese, S. P. Kulik, E.v. Moreva, A. p shurupov, Phys. Rev. Lett. 105, 010404 (2010).

\bibitem{Prosen} T. Gorin, T. Prosen, H. Seligman, M. Znidaric, Phys. Rep. 435, 33-156 (2006).

\bibitem{SJGu} S.-J. Gu, Int. J. Mod. Phys. B 24, 4371 (2010).
%

\bibitem{Langford} A. Gilchrist, N. K. Langford, M. A. Nielsen, Phys. Rev. A 71, 062310 (2005).



\bibitem{RMRS} R. Muthuganesan,  R. Sankaranrayanan, Phys. Lett. A 381, 3028-3032 (2017).





\bibitem{Marian} P. Marian, T. A. Marian, Phys. Rev. A 86, 022340 (2012).




\bibitem{Milburn} G. J. Milburn, J. Phys. A 17,  737-745  (1984).

\bibitem{WRPT} W. P. Bowen, R. Schnabel, P. K. Lam,  T. C. Ralph, Phys. Rev. A 69, 012304 (2004).




\end{thebibliography}
\end{document}